\documentclass[pdflatex,sn-mathphys-num]{sn-jnl}

\usepackage{graphicx}
\usepackage{multirow}%
\usepackage{amsmath,amssymb,amsfonts}%
\usepackage{amsthm}%
\usepackage{mathrsfs}%
\usepackage[title]{appendix}%
\usepackage{xcolor}%
\usepackage{textcomp}%
\usepackage{manyfoot}%
\usepackage{booktabs}%
\usepackage{algorithm}%
\usepackage{algorithmicx}%
\usepackage{algpseudocode}%
\usepackage{listings}%
\usepackage{lineno}

\theoremstyle{thmstyleone}%

\theoremstyle{thmstyletwo}%

\theoremstyle{thmstylethree}%

\begin{document}

\title[Comparing surface and deep horizontal distributions of depth-keeping particles in shallow fluid layers]{Comparing surface and deep horizontal distributions of depth-keeping particles in shallow fluid layers}

\author{\fnm{Lenin Mois\'es} \sur{Flores Ram\'irez}}\email{l.m.flores.ramirez@tue.nl}

\author{\fnm{Matias} \sur{Duran-Matute}}\email{m.duran.matute@tue.nl}

\author*[~]{\fnm{Herman J. H.} \sur{Clercx}}\email{h.j.h.clercx@tue.nl}

\affil{\orgdiv{Fluids and Flows group and J.M. Burgers Center for Fluid Mechanics, Department of Applied Physics and Science Education}, \orgname{Eindhoven University of Technology}, \orgaddress{\street{P.O. Box 513}, \city{Eindhoven}, \postcode{5600 MB}, \state{
Noord-Brabant}, \country{The Netherlands}}}

\abstract{This study examines whether the dispersion of passive particles at the free surface of a generic (nonturbulent) shallow flow can reliably represent the behavior of depth-keeping particles below the surface. A shallow configuration characterize many aquatic environments, such as coastal regions and lakes, where horizontal scales far exceed vertical ones, large-scale flow structures dominate, and observations are sometimes limited to the surface. We compare surface and subsurface horizontal velocities in both direction and magnitude, identifying distinct behaviors depending on the parameter $Re_F\delta^2$, where $Re_F$ is the Reynolds number based on forcing, and $\delta$ is the aspect ratio between the fluid layer depth and the horizontal forcing scale. At low $Re_F\delta^2$, deep flows match the surface flow in direction throughout the layer, but not in magnitude. At high $Re_F\delta^2$, the magnitude matches (outside the bottom boundary layer), but not always the direction. Despite these differences, for all $Re_F\delta^2$, surface particle patterns correlate with those in the upper quarter of the fluid layer. Filamentary structures caused by horizontal flow convergence remain spatially aligned within this region. Below it, at intermediate $Re_F\delta^2$, deep filaments become diffuse and eventually vanish. At high $Re_F\delta^2$, filaments persist at depth, but become spatially misaligned with surface filaments. These findings suggest that in shallow environments, surface observations can quantitatively infer subsurface transport processes in the upper quarter of the fluid layer. For the deeper part, knowledge of the vertical profiles of the mean flow yields insights into the horizontal transport processes.

}

\keywords{Shallow flow, horizontal dispersion, subsurface dispersive processes, correlation dimension}

\maketitle

\section*{Article highlights}
\begin{itemize}
    \item The surface particle patterns reflect {\color{black}subsurface features from} the top quarter of the shallow fluid layer
    \item Four characteristic regimes {\color{black}are identified for horizontal particle distributions as function of height} and flow forcing
    \item Accurate {\color{black}inference of subsurface dispersion} requires knowledge of vertical variation of the horizontal flow field
\end{itemize}

\section{Introduction}\label{sec1}

Drifters are devices used to track water currents, making them ideal for studying dispersion, an inherent Lagrangian quantity \cite{Lumpkin2017AdvancesDrifters}. Surface drifters, in particular, are specifically designed to follow currents in the upper layer of the water column. They have been used to investigate, for example, the fundamental properties of ocean dispersion \cite{Corrado2017}, oil spills \cite{Novelli2017AScale}, spread of radionuclides \cite{Rypina2014DrifterbasedRadionuclides}, and the evolution of garbage patches \cite{Maximenko2012PathwaysDrifters,vanSebille2015ADebris}. However, these drifters provide only surface information by their very nature, restricting insight into deeper horizontal dispersive processes. 

To overcome this limitation, several strategies have been proposed, including the deployment of subsurface drifters \cite{Hancock2022DispersionMexico,Meunier2021AMexico,Rypina2021ObservingDepths}, though with temporal and spatial constraints, and the use of numerical models \cite{Zhong2013SubmesoscaleMexico,Zhong2017ObservedSea}. Another approach involves inferring deep currents from surface ones \cite{Wang2013ReconstructingData}, which are more readily accessible through satellite observations that offer fairly high spatial and temporal coverage \cite{Rohrs2023SurfaceMethods}. Once reconstructed, these subsurface currents can be used to simulate virtual drifters and analyze their dispersive behavior. 

Nevertheless, when only surface information is available, one may question to what extent surface dispersion can represent the dispersion occurring below the surface. In the ocean mixing layer, \citet{Berti2021LagrangianInstabilities} found that the surface and deep horizontal dispersion of (depth-keeping) particles is largely controlled by large-scale vertical shear of horizontal velocities. They suggested that if large-scale shear is known, it should be possible to infer how a tracer at depth separates from one at the surface. Furthermore, large-scale vertical shear is reported to be effective in {\color{black}enhancing} the horizontal relative dispersion of neutrally buoyant particles at different depths, often more so than turbulent motions alone, due to persistent velocity differences experienced by the particles \cite{Lanotte2016EffectsDispersion,Liang2018HorizontalLayer}. However, the similarities between surface and deep dispersion are influenced by hydrodynamic conditions. For example, during certain months in the Gulf of Mexico, the dispersion of depth-keeping particles located one meter below the surface (within the mixed layer) does not correlate with that at a depth of one hundred meters. Specifically, near the surface, particles are observed to form spiral structures around eddies, while those at greater depth remain uniformly distributed \cite{Zhong2012Field}. In contrast, in other months, similar spiral structures appear at both depths \cite{Zhong2013SubmesoscaleMexico}. Moreover, this discrepancy might become more pronounced when considering particles that do not maintain a fixed depth. If particles at depth are able to move vertically, due, for example, to turbulent fluctuations, this can lead to distinct horizontal dispersion patterns and consequently to a different fate for the particles, as observed in a small bay \cite{Jalon-Rojas2019TechnicalMicroplastics}.

{\color{black}By depth-keeping particles, we refer to particles that are capable of maintaining a relatively constant depth within a fluid column. In numerical studies such as those mentioned above \citep{Berti2021LagrangianInstabilities,Lanotte2016EffectsDispersion,Liang2018HorizontalLayer,Zhong2012Field,Jalon-Rojas2019TechnicalMicroplastics}, this behavior is typically achieved by imposing a zero vertical velocity on the particles. Although this may appear as a numerical idealization, depth regulation to counteract upwelling or downwelling currents occurs naturally and is technologically achievable. This situation occurs with small aquatic organisms that remain in a thin horizontal layer below the free surface of lakes or oceans, for example, by swimming or by active regulation of their buoyancy; see Refs. \cite{Franks1992SinkFronts,FloresRamirez2025VerticalFlows}. Certain species of zooplankton actively swim to remain within preferred depth layers \citep{Genin2005SwimmingAggregation,Chen2021DielPatchiness}. Miniature underwater robotic drifters employ controllable buoyancy mechanisms to maintain a prescribed depth \citep{Jaffe2017ADynamics,Morgan2021}. These examples illustrate that depth-keeping behavior is realistic and relevant and motivate our focus on how such particles experience horizontal dispersion in shallow flows.}

{\color{black}This study examines the connection between surface and deep horizontal dispersion in an idealized, continuously forced (nonturbulent) shallow flow using Lagrangian statistical measures, such as particle-pair correlations and associated correlation dimension and the vertical correlation of the location of particle patches, as a function of depth.} The shallow configuration reflects common natural environments, such as coastal ocean regions and lakes, {\color{black}where the shallow flow is fully turbulent on small scales, produced by ejection and sweep events near the bottom, yet its large-scale motion frequently exhibits two-dimensional (2D) features \citep{Jirka2001}, and where relevant physical and biological processes usually occur in the upper part of the water column. A defining feature of strictly 2D flows is their tendency to self-organize, from an initial disorganized state, into one with large-scale coherent vortices; see, for instance, Ref. \cite{McWilliams1984}. Such vortices appear in shallow environments due to the strong disparity between horizontal and vertical dimensions; horizontal extents typically extend kilometers, while depths are often only a few meters \citep{Clercx2022Quasi-2DLayers}. However, as shown by laboratory experiments with shallow fluid layers and numerical simulations of shallow fluid flows \cite{VanHeijst2014}, such flows also contain secondary vertical and horizontal motions superimposed onto those large-scale vortices (the primary flow). For this reason, we refer to the flow as quasi-2D when horizontal primary motions dominate but secondary motions, although present, remain weak, and we consider it to exhibit a more three-dimensional (3D) character when secondary motions become dynamically relevant and lead to significant variation across the depth. The relative strength of these secondary motions thus determines the degree to which the flow departs from quasi-2D behavior.} The present study focuses on the impact of these primary and secondary large-scale motions on horizontal dispersion, and {\color{black}we mostly exclude} the impact of small-scale turbulence on dispersion.

As an example to illustrate the transition between the quasi-2D and 3D state, we consider shallow dipolar vortices that decay freely \cite{Duran-Matute2010DynamicsVortices}. The strength of horizontal and vertical secondary flows in these shallow dipoles scales with $Re\delta^2$ and $Re\delta^3$, respectively. Here, $Re=UR/\nu$ is the Reynolds number, $\delta=H/R$ is the aspect ratio, with $U$ a characteristic velocity (the dipole speed), $R$ the radius of the dipole, $H$ the depth of the fluid and $\nu$ the kinematic viscosity. When $Re\delta^2 \lesssim 6$, the flow remains predominantly horizontal (quasi-2D), as secondary motions are suppressed by viscous damping from bottom friction, and the vortex structure remains vertically coherent. The flow can be approximated as the flow taken at the surface modulated with a Poiseuille-like vertical dependence to accommodate the no-slip bottom {\color{black}boundary \cite{Juttner1997,Paret1997,Akkermans2008}.} For $Re\delta^2\gtrsim15$, stronger secondary motions lead to a breakdown of the vertical coherence of the dipole, and the magnitude and direction of the flow strongly depend on the vertical position in the fluid layer. In the case of a continuously forced dipole, the transition occurs at $Re\delta^2=\pi^2$ \cite{Duran-Matute2011}. 

Similar findings were obtained for a lattice of shallow forced vortices \cite{FloresRamirez2025AsymmetricFlows}. When $Re_F\delta^2\lesssim10$, where $Re_F$ is the Reynolds number based on forcing and $\delta$ is the aspect ratio between the depth of the fluid layer and the horizontal forcing length scale, the vertical variation of the horizontal velocity field is characterized by a Poiseuille-like profile. On the other hand, for $Re_F\delta^2\gtrsim 10$, the flow gradually becomes unsteady and spatially {\color{black}disorganized, with a vertical structure comprising an inviscid interior and a thin bottom boundary layer, and floating particles on the surface form filaments, as also earlier observed by  \citet{Akkermans2012ArraysDispersion}. Despite the quasi-2D nature of the 3D flow, the horizontal surface flow is not divergence-free, even with extremely weak secondary motions. As a result, passive particles floating on the surface, although initially distributed homogeneously, eventually accumulate into thin, elongated filaments in regions of convergent flow, i.e., where the horizontal divergence is negative. This behavior is markedly different from the strictly 2D case, where the flow is divergence-free and, consequently, the particles remain evenly distributed
\cite{Akkermans2012ArraysDispersion}.} 

{\color{black}By tracking tracer particles on the rigid free surface and depth-keeping tracers at various depths in the fluid layer, we provide input for a variety of Lagrangian statistical measures to determine the extent to which surface observations can statistically represent those at depth, or, in other words, if these observations can serve as a qualitative (statistical) predictor of the particle distributions below the surface. This approach implies that no instantaneous comparisons between surface and deeper particle trajectories are made.} Intuitively, surface dispersion may seem a reliable proxy for deep {\color{black}horizontal} dispersion when the shallow flow behaves quasi-2D, since the surface flow is vertically coherent {\color{black}throughout the fluid layer, with a Poiseuille-like profile of the horizontal velocity magnitude in the vertical direction \citep{Juttner1997,Paret1997,Akkermans2008}. When} the flow structure becomes 3D, it is expected that the surface dispersion is no longer representative of the deep {\color{black}horizontal} dispersion. However, contrary to intuition, we will show that the relation between surface and deep {\color{black}horizontal} dispersion results is nontrivial because of the tendency of depth-keeping particles to accumulate in certain regions of the flow due to the underlying flow topology and secondary flows. 

The paper is organized as follows. In Sect.~\ref{sec:statement_problem}, the problem is briefly formulated and the numerical methods are described. Section \ref{sec:flow_comparison} is devoted to the comparison of surface flow with deeper horizontal flow fields from a Eulerian perspective. Section \ref{sec:particle_comparison} focuses on comparing the particle distributions on the surface with those at depth, and four regimes are distinguished that characterize the relation between surface and deep horizontal dispersion. In that section, a first exploration of the impact of vertical particle excursions and small-scale turbulence on horizontal particle transport is also presented. Finally, in Sect.~\ref{sec:conclusion}, the main conclusions of this study are summarized.

\section{Problem definition and numerical methods} \label{sec:statement_problem}

We investigate the horizontal transport of particles embedded in a shallow fluid layer of horizontal extent $L$ and depth $H$, with $H\ll L$. Transport at the stress-free, rigid surface of the layer is compared with that at different depths,  from below the surface to near the no-slip bottom. The particles on the surface are advected only by the horizontal flow since the vertical velocity is zero at the rigid surface. {\color{black}Similarly, particles at deeper depths are advected by the horizontal flow alone by artificially setting their vertical velocities to zero.} The flow is continuously forced horizontally with a characteristic length scale $L_f>H$, achieving a statistically steady state. Further details on the implementation are given below.

\subsection{Numerical setup} 

Consider a homogeneous, incompressible fluid whose motion is governed by the Navier-Stokes equations and the continuity equation. Expressed in dimensionless form, these equations are
 
\begin{equation} 
\frac{\partial\mathbf{u}}{\partial t}+(\mathbf{u}\cdot\boldsymbol{\nabla})\mathbf{u}=-\boldsymbol{\nabla} p+\dfrac{1}{Re_F}\nabla^2\mathbf{u}+\mathbf{f}, \label{eq:ns_eq}
\end{equation}

\begin{equation}
\boldsymbol{\nabla}\cdot\mathbf{u}=0, \label{eq:mass_conservation}
\end{equation}
with $\textbf{u} = (u,v,w) $ the flow velocity, $t$ the time, $p$ the pressure, and $\textbf{f}$ an external body force. {\color{black}All spatial coordinates are non-dimensionalised by the forcing scale $L_f$, and the fluid velocity, time, and pressure by the velocity scale $U=\sqrt{FL_f/\rho}$ (with $F$ the forcing magnitude per unit volume and $\rho$ the fluid density), the time scale $L_f/U$, and $\rho U^2$, respectively.} The Reynolds number is defined as $Re_F=(L_f/\nu)\sqrt{FL_f/\rho}$, with $\nu$ the kinematic viscosity. {\color{black}The gravitational force (in the vertical direction) was intentionally neglected; therefore, no Froude number appears in Eq. \eqref{eq:ns_eq}.} The fluid motion is described in Cartesian coordinates $\mathbf{x}=(x,y,z)$, where the horizontal movement occurs in the $xy$-plane, while $z$ is the vertical direction.

The flow is driven by a fixed-amplitude steady force $\textbf{f} = \left ( \partial q/\partial y, -\partial q/\partial x,0 \right )$, with $q=\sin(\pi x)\sin(\pi y)/(2\pi^2)$. This force only acts in the horizontal direction and on a single scale, which is equal to unity. The force does not have a vertical dependence and satisfies ${\boldsymbol\nabla} \cdot \textbf{f} = 0$. The length of the domain determines the total number of induced vortices, so that $L$ must be an integer multiple of $L_f$. {\color{black}For further details, see Ref. \citep{FloresRamirez2025AsymmetricFlows}.}

\begin{figure}
    \centering
    \includegraphics[scale=.99]{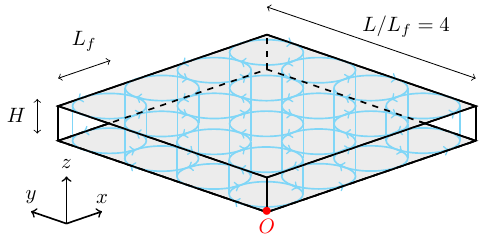}
    \caption{Computational domain for the 3D simulations. The no-slip bottom wall and stress-free (rigid) upper boundary are colored in gray. The cyan lines sketch how the forcing is acting on the fluid. The red dot indicates the origin $O$ of our reference frame.}
    \label{fig-box}
\end{figure}

Numerical simulations based on Eqs. \eqref{eq:ns_eq} and \eqref{eq:mass_conservation} are performed using a finite element code \cite{COMSOLMultiphysicsIntroductionMultiphysics}. The computational domain is a rectangular box of dimensions $4 \times 4 \times \delta$ with $\delta = H/L_f$, see Fig.~\ref{fig-box}. A rigid and stress-free boundary condition is applied to the surface (top) so that free surface deformations are neglected. At the bottom, a no-slip boundary condition is applied. The lateral boundaries are periodic, which implies that $L/L_f$ is restricted to even integer values. The parameter space is explored by performing simulations for three values of the aspect ratio, i.e. $\delta=0.1,\, 0.3$ and $0.5$, and different values of $Re_F$, as shown in Table \ref{table:param}. Most of these flow configurations were used in a previous study that focused on Eulerian flow statistics~\cite{FloresRamirez2025AsymmetricFlows}, but for the current investigation more simulations were performed and have been run longer.   

The domain is discretized using an unstructured triangular mesh in the horizontal and a structured rectangular mesh in the vertical, with finer elements near the top and bottom boundaries to resolve the vertical gradients in the flow field. The number of mesh elements changes depending on the value of $Re_F$ and $\delta$ to achieve accurate results with the least computational cost. The resolutions range from approximately $40,000$ to $700,000$ elements, with 20 to 30 elements in the vertical direction. The simulation accuracy was evaluated by the mean absolute divergence, since the zero divergence condition (Eq.~\eqref{eq:mass_conservation}) is sensitive to insufficient resolution. For most of our simulations, it was approximately 1-2\% of the root-mean-square (rms) vorticity, which was considered sufficient. In addition, grid convergence was checked by comparing the total energy of the flow with different mesh refinements. The difference in energy between fine and coarser meshes is always less than 0.5-1\%, which we consider sufficient. Time stepping uses backward differentiation formulas of variable order and variable step size, with order and step size determined by the numerical code based on the simulation requirements \cite{COMSOLMultiphysicsIntroductionMultiphysics}.

\begin{table}
\centering
\caption{\small{Overview of the control parameters $\delta$ and $Re_F$.}}
\label{table:param}
\begin{tabular}{@{}ll@{}}  
\toprule
$\delta$ & $Re_F$ \\
\midrule
0.1 & 200, 300, 475, 600, 1000, 1650, 2825, 4000, 6000 \\
0.3 & 22, 44, 70, 115, 220, 325, 430, 745, 1060   \\
0.5 & 18, 35, 50, 70, 90, 130, 175, 290, 400 \\
\botrule
\end{tabular}
\end{table}

\subsection{Lagrangian particle tracking} \label{sec:lagrangian_track}

The trajectories of $N_p$ particles are obtained by integrating the equations of motion of each particle $p$:
\begin{equation}
\dfrac{d\textbf{x}^p(t)} {dt}=\mathbf{u}^p_f, \label{eq:eq_motion_particles} 
\end{equation}
where $\mathbf{u}^p_f=(u^p_f,v^p_f,0)$ is the horizontal flow velocity at the position of the particle $\textbf{x}^p=(x^p,y^p,z^p)$. The time integration of Eq.~\eqref{eq:eq_motion_particles} is performed using a fourth-order Runge-Kutta method, and the fluid velocity at the position of each particle is calculated using linear interpolation in space. 

On the surface (at $z=\delta$), $N_p = 10^5$ randomly distributed particles are released in a square array that spans $0.05\leq x^p\leq 3.95$ and $0.05\le y^p \leq 3.95$ at $t=0$. Periodic boundary conditions are applied to the positions $(x^p,y^p)$ when the particles cross the horizontal boundaries. This configuration is replicated vertically at nine different depths (equidistant) within the range $0.1\delta \leq z \leq 0.9\delta$. The particles are released at these depths and on the surface once the flow reaches a statistically steady state. 

The particles on the surface remain confined to it, following only the horizontal flow. Particles within the shallow fluid layer are restricted to their release depth, passively moving by horizontal flow at that depth. This setup is analogous to drifters deployed at various depths in oceanic waters \cite{Meunier2021AMexico,Rypina2021ObservingDepths} or the limiting case of passive particles constrained vertically \cite{FloresRamirez2025VerticalFlows,DePietro2015ClusteringTurbulence}.

\section{Results}\label{sec2}

\subsection{Surface flow versus deep flows} \label{sec:flow_comparison}

This section focuses on comparing the horizontal flow at the free surface with the horizontal motion at a different depth. To achieve this, we adopt the statistical tools proposed by \citet{Martell2019}. In their study, they generated a flow by electromagnetically forcing an immiscible two-layer fluid and compared the flow at the free surface with the motion at the fluid interface. In contrast, the simulated flows in our study are not stratified, and we will compare the surface flow with the flow throughout the entire layer with emphasis on the speed ratio and alignment of the horizontal velocity vectors.

{\color{black}Eulerian characteristics of the present and similar forced (and decaying) shallow flows have been previously reported \cite{Duran-Matute-JFM2010,Duran-Matute2011,FloresRamirez2025AsymmetricFlows}.} The features relevant to this work are the following. The flow regimes of the simulations can be characterized by the parameter $Re_F\delta^2$ provided $\delta\ll 1$. For small $Re_F\delta^2$, bottom friction dominates over inertia and the flow is in the viscous regime. The flow exhibits a steady checkerboard vortex pattern as a linear response to the forcing, and a Poiseuille-like vertical profile of the horizontal velocity components due to the dominance of viscous forces. However, with increasing $Re_F\delta^2$, when the inertial and viscous forces are comparable, the flow becomes unsteady. For even larger $Re_F \delta^2$, inertia becomes comparable to the external forcing, and the flow is in the inertial regime. It becomes more disorganized and consists of an inviscid interior and a thin bottom boundary layer with {\color{black}thickness $z_b$}. The transition between the viscous and inertial regimes occurs for $Re_F \delta^2 \approx 10$~\cite{FloresRamirez2025AsymmetricFlows}.

\subsubsection{Speed ratio}

We examine the relationship between the horizontal flow speed at a given depth $z$ and the speed at the free surface ($z=\delta$):

\begin{equation}
 s=\dfrac{|| \mathbf{u}_h(x,y,z,t) ||}{|| \mathbf{u}_h(x,y,\delta,t)||}=\dfrac{u_h}{u_h|_{z=\delta}},    
\end{equation}
\noindent where $\mathbf{u}_h=(u,v)$ denotes the horizontal velocity vector and $||\cdot ||$ is the norm of a vector. This ratio expresses how the magnitude of the flow speed at depth compares to the surface speed.

\begin{figure}
    \centering
    \includegraphics[scale=.45135]{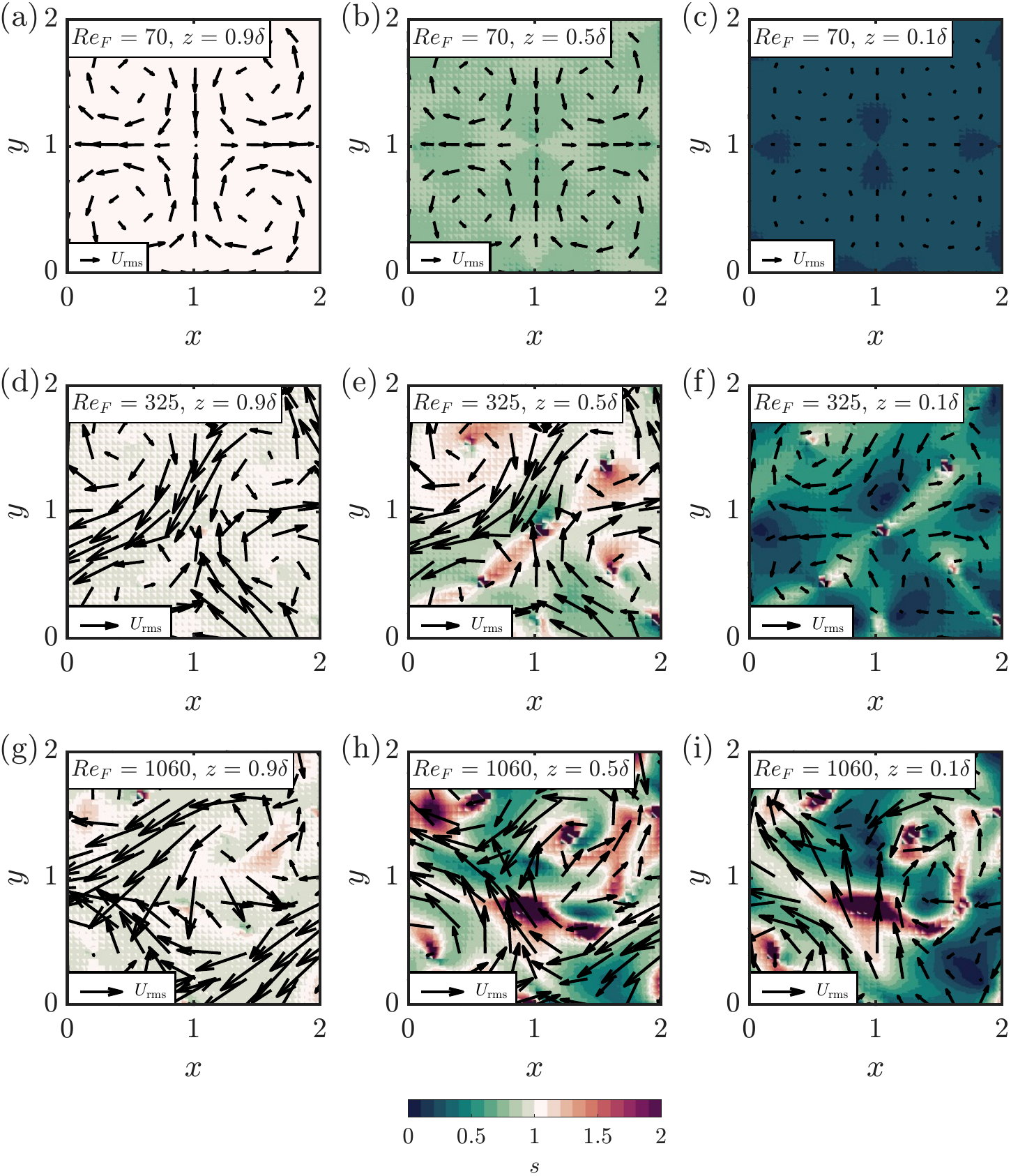}
    \caption{Instantaneous fields of the speed ratio $s$ at $z=0.9\delta$, $0.5\delta$ and $0.1\delta$ (from left to right) for flows with $Re_F=70$, $325$, and $1060$ (from top to bottom), as indicated by the label in each panel. For all flows, $\delta=0.3$. The arrows indicate the horizontal velocity fields at the corresponding $z/\delta$ planes. For simplicity, only one quarter of {\color{black}the horizontal domain is shown. Snapshots of the vertical velocity field in the full (horizontal) domain for the three cases shown in panels (b), (e) and (h) can be found in Fig. 7 of Ref.~\cite{FloresRamirez2025AsymmetricFlows}.}}
    \label{fig:sm1_fields_d03}
\end{figure}

Figure \ref{fig:sm1_fields_d03} shows some snapshots of the speed ratio $s$ for $z/\delta=0.9$, 0.5, and 0.1, taken from three flows with $\delta=0.3$. For $Re_F=70$ ($Re_F\delta^2=6.3$, Fig.~\ref{fig:sm1_fields_d03}a-c), the fields obtained for the speed ratios are horizontally uniform and vary only with depth: from $s\approx 1$ at $z=0.9\delta$, decreasing to $s\approx 0.8$ at $z=0.5\delta$, and reaching $s\approx0.2$ at $z=0.1\delta$.  Similarly, for $Re_F=325$ ($Re_F\delta^2=29.3$, Fig.~\ref{fig:sm1_fields_d03}d) and $Re_F=1060$ ($Re_F\delta^2=95.4$, Fig.~\ref{fig:sm1_fields_d03}g), $s\approx 1$ almost everywhere at $z=0.9\delta$, indicating that the velocity magnitudes at this depth are comparable to those at the surface. However, for these larger $Re_F$ values, regions where $s>1$ are observed at $z=0.5\delta$ (see Figs.~\ref{fig:sm1_fields_d03}e and h), indicating velocity magnitudes at lower depths that exceed surface speeds. These regions are more prominent and extend to $z=0.1\delta$ for $Re_F=1060$, representing flows dominated by strong inertia (Fig.~\ref{fig:sm1_fields_d03}i). This is not observed for $Re_F=325$, where $s$ decreases below one (see Fig.~\ref{fig:sm1_fields_d03}f), because $z=0.1\delta$ is clearly within the viscous boundary layer where velocity magnitudes are suppressed. 

\begin{figure}
    \centering
    \includegraphics[scale=.2666666]{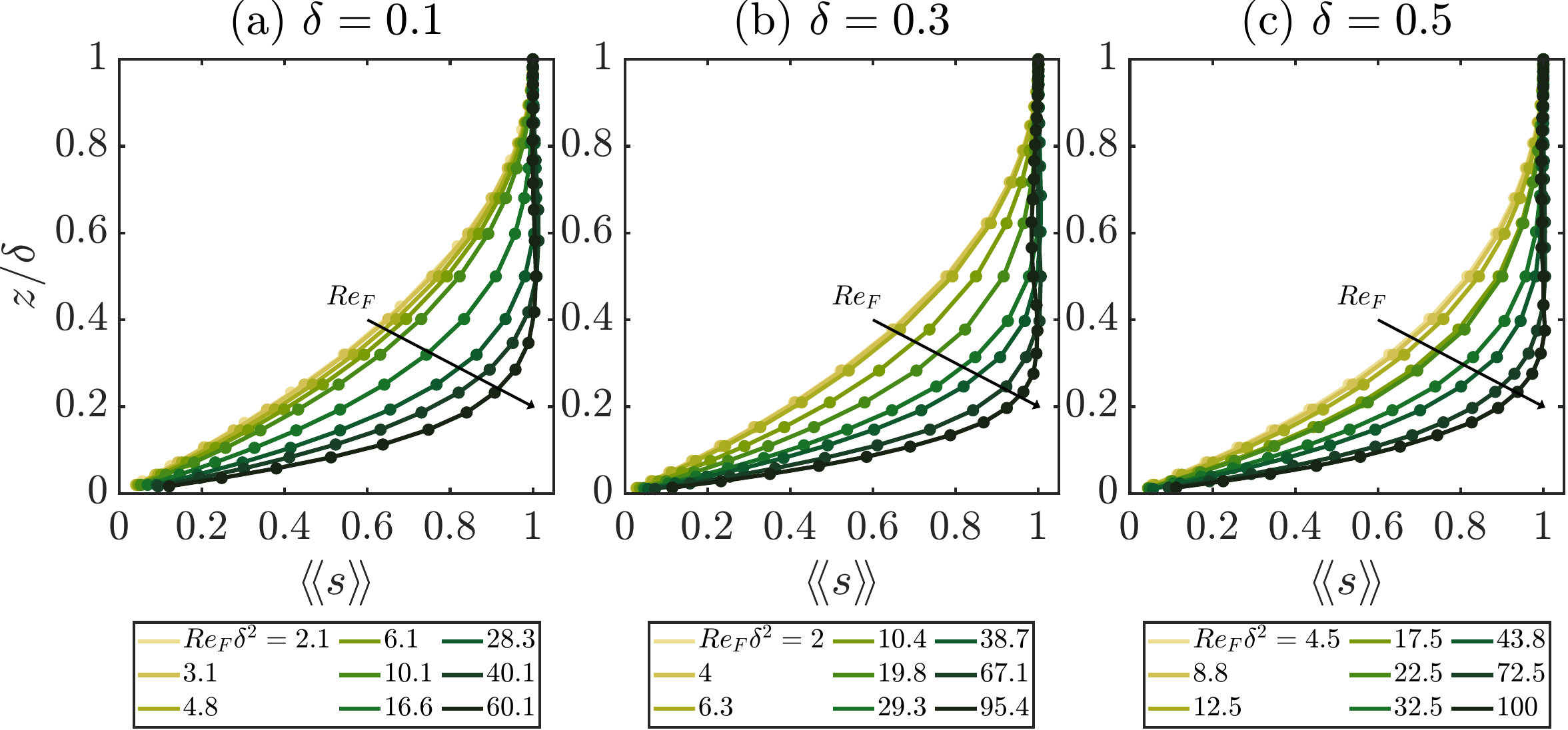}
    \caption{Vertical profiles of the average speed ratio $\langle\!\langle s \rangle\!\rangle$ for all simulations. Each subfigure corresponds to a different $\delta$ value as indicated by their titles. The arrow in the panels indicates increasing values of $Re_F$.}
    \label{fig:sm1_profiles}
\end{figure}

To examine the vertical dependence of the speed ratio $s$ in all simulations, we focus on the vertical profiles of the mean value of $s$, which is obtained by averaging over both the horizontal directions and time (starting when the flow is statistically steady) \cite{FloresRamirez2025AsymmetricFlows} and is denoted $\langle\!\langle s \rangle\!\rangle$. The results are shown in Fig.~\ref{fig:sm1_profiles}. To mitigate the influence of spurious and excessively high values of $s$ on the mean, the average is calculated from data within the first 95\% of the $s$ distribution. 

The profiles for the smaller $Re_F\delta^2$ values show that the ratio of velocity magnitudes decreases monotonically with depth (relative to the surface speed), closely resembling a Poiseuille profile. This occurs because of the dominance of vertical viscous diffusion. As $Re_F\delta^2$ increases, the vertical profiles of $\langle\!\langle s\rangle\!\rangle$ depart from this Poiseuille-like profile. For higher $Re_F\delta^2$ values, there is a portion of the profile in which $\langle\!\langle s\rangle\!\rangle$ remains nearly constant, followed by a rapid decrease to zero near the bottom reflecting the fact that these flows are composed of an almost inviscid interior and a boundary layer close to the bottom. 

\begin{figure}
    \centering
    \includegraphics[scale=0.4]{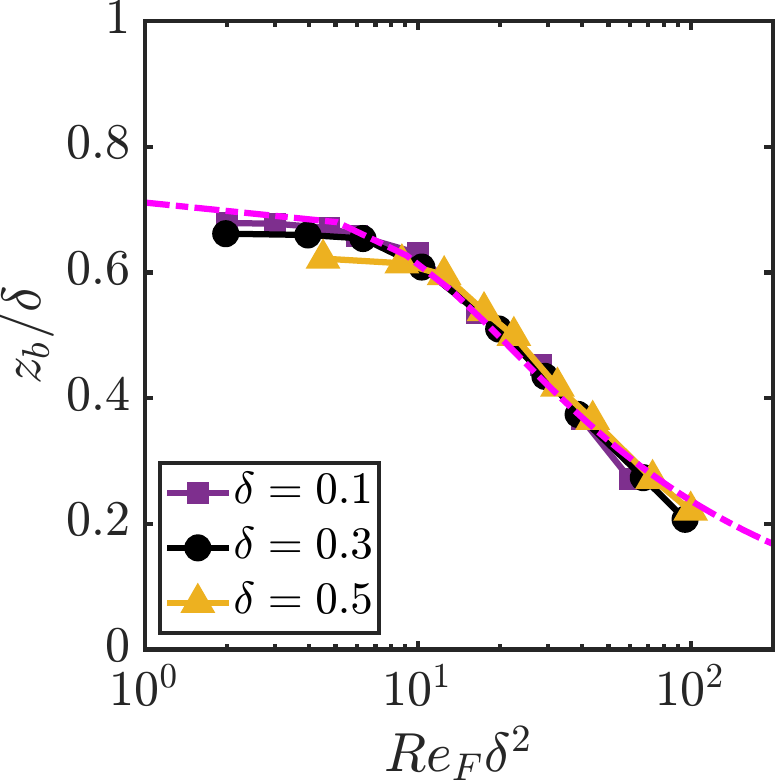}
    \caption{Normalized boundary layer thickness $z_b/\delta$ versus $Re_F\delta^2$. The dashed line corresponds to the function  $z_b/\delta=0.7[1+(Re_F\delta^2/b)^2]^{-1/4}$, see Eq.~(\ref{eq:zbd}), with $b=11.1\pm 0.8$ a fitting parameter.}
    \label{fig:zb_vs_ReFD2}
\end{figure}

The thickness of this boundary layer can be determined from the vertical profiles of rms velocity, $U_\mathrm{rms}=\langle\!\langle u_h^2 \rangle\!\rangle^{1/2}$, and is defined as the depth $z_b$ at which $U_\mathrm{rms}$ reaches 90\% of its surface value, that is, $U_\mathrm{rms}(z_b)=0.9U_\mathrm{rms}(\delta)$. In Fig.~\ref{fig:zb_vs_ReFD2} it is shown that the values of $z_b/\delta$ collapse for each $\delta$ considered on a single curve when plotted against the parameter $Re_F\delta^2$. In the viscous regime ($Re_F\delta^2\lesssim10$), we expect $z_b/\delta=2\arcsin(0.90)/\pi\approx 0.7$. This corresponds to the value of $z_b/\delta$ for a Poiseuille velocity profile $U_\mathrm{rms}(z)/U_\mathrm{rms}(\delta)=\sin(\pi z/2\delta)$.  In the inertial regime ($Re_F\delta^2\gtrsim 10$), $z_b/\delta$ is expected to scale as $(Re_F\delta^2)^{-1/2}$. {\color{black}This scaling is derived from a balance between inertia, $(\mathbf{u}\cdot\boldsymbol{\nabla})\mathbf{u}\sim U^2/L_f$, and viscous forces, $\nu{\boldsymbol{\nabla}}^2 {\mathbf{u}}\sim \nu\partial^2 {\mathbf{u}}/\partial z^2 \sim \nu U/z_b^2$ \cite{Duran-Matute2011}. Here, it is assumed that the velocity will vary vertically with a length scale equivalent to the thickness of the boundary layer.} Based on these expectations, we fit the function 
\begin{equation}
\frac{z_b}{\delta}=0.7\left[1+\left(\frac{Re_F\delta^2}{b}\right)^2\right]^{-1/4}\label{eq:zbd}
\end{equation}
to the data, where $b=11.1\pm 0.8$ is a fitting parameter. Interestingly, $b$ is relatively close to $Re_F \delta^2 = 10$, which is the value at which the transition from viscous to inertial regime occurs. This function approaches the expected values for the viscous (inertial) regime as $Re_F\delta^2$ becomes smaller (larger) than $b$. We observe that the values of $z_b/\delta$ follow this function reasonably well. {\color{black}As expected, $z_b$ decreases because the viscous bottom region narrows (or, equivalently, the inviscid interior expands) as the flow transitions from the viscous to the inertial regime. As we show later (see Section~\ref{Sec4.5}), this depth serves not only as the boundary between the viscous and inviscid regions, but also separates two distinct horizontal particle dispersion behaviors above and below $z_b$.}

\subsubsection{Alignment between flows}\label{sec-cos}

The angle $\theta$ between the flow
directions at the free surface and at depth $z$ is given by the following expression,

\begin{equation}
    \cos{\theta}=\dfrac{\mathbf{u}_h(x,y,\delta,t)\cdot \mathbf{u}_h(x,y,z,t)}{||\mathbf{u}_h(x,y,\delta,t) ||\, ||\mathbf{u}_h(x,y,z,t) ||}, \label{eq:cos_theta}
\end{equation}

\noindent and quantifies the alignment of the horizontal velocity vectors at the surface and at depth $z$. When $\cos\theta=1$ (i.e. $\theta=0^\circ$), they are aligned, pointing in the same direction, and anti-aligned for $\cos\theta=-1$  (i.e. $\theta=180^\circ$). 

\begin{figure}[h!]
    \centering
    \includegraphics[scale=.5015]{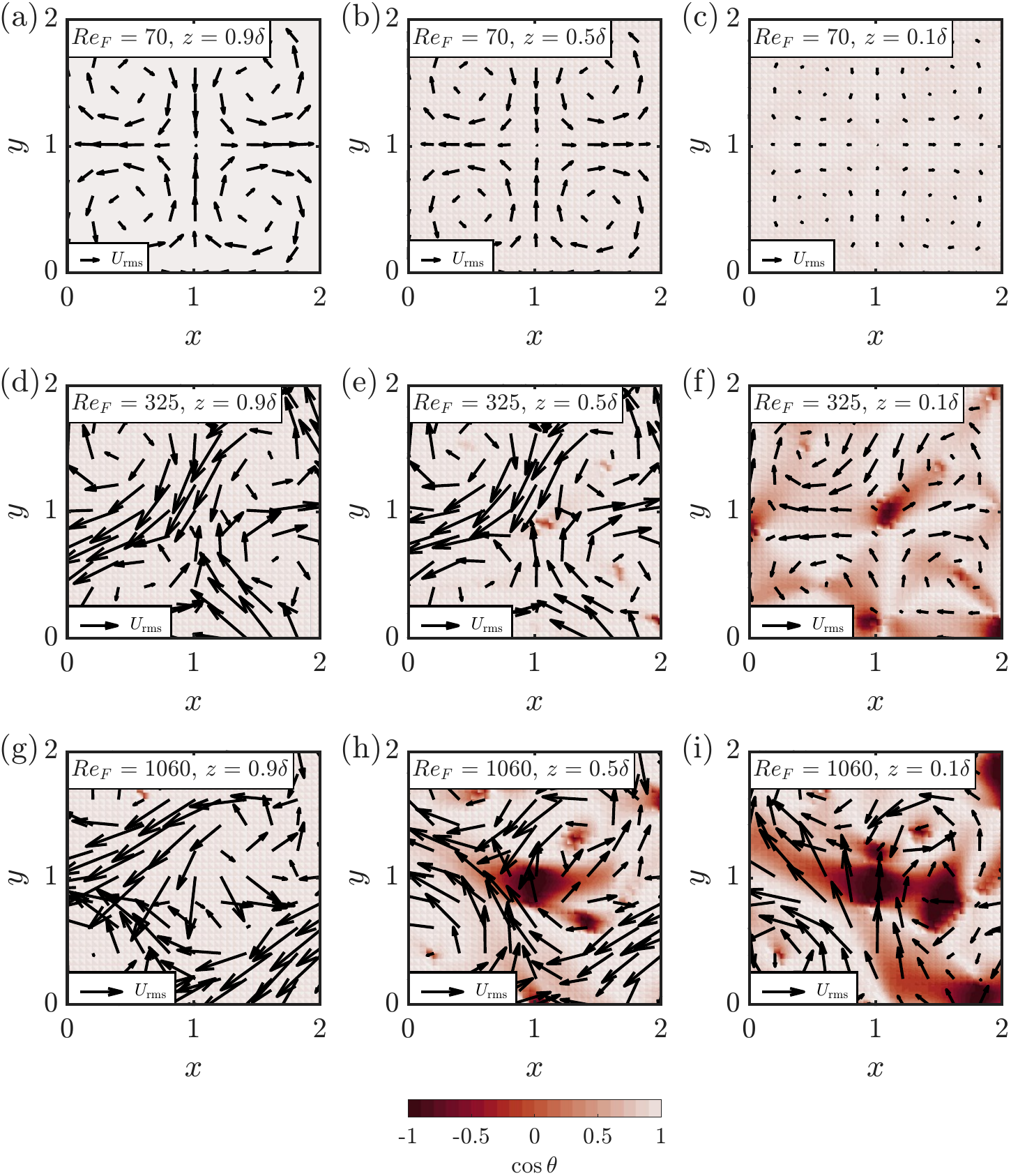}
    \caption{Instantaneous fields of $\cos\theta$ at $z=0.9\delta$, $0.5\delta$ and $0.1\delta$ (from left to right) for flows with $Re_F=70$, $325$, and $1060$ (from top to bottom), as indicated by the label in each panel. For all flows, $\delta=0.3$. The arrows indicate the horizontal velocity fields at the corresponding $z/\delta$ planes. For simplicity, only one quarter of the domain is shown.}
    \label{fig:cos0_fields_d03}
\end{figure}

Figure \ref{fig:cos0_fields_d03} presents snapshots of $\cos\theta$ taken from the same flows and at the same depths as in Fig.~\ref{fig:sm1_fields_d03}. Due to its proximity to the surface, the horizontal flow at $z=0.9\delta$ is perfectly aligned with that at the surface almost everywhere for all values of $Re_F$ (Figs.~\ref{fig:cos0_fields_d03}a, d and g). For $Re_F=70$ ($Re_F\delta^2=6.3$), flows at $z=0.5\delta$ and $z=0.1\delta$ (Figs.~\ref{fig:cos0_fields_d03}b and c, respectively) remain aligned with the surface flow. For $Re_F=325$ ($Re_F\delta^2=29.3$) and $z=0.5\delta$ (Fig.~\ref{fig:cos0_fields_d03}e), few anti-alignment values are observed in the center of vortices and the stagnation points. Anti-alignment values become more frequent at $z=0.1\delta$ (Fig.~\ref{fig:cos0_fields_d03}f). At $Re_F=1060$ ($Re_F\delta^2=95.4$), anti-alignment values become more dominant for $z\lesssim 0.5\delta$ (Figs.~\ref{fig:cos0_fields_d03}h and i). 

\begin{figure}
    \centering
    \includegraphics[scale=.26666666]{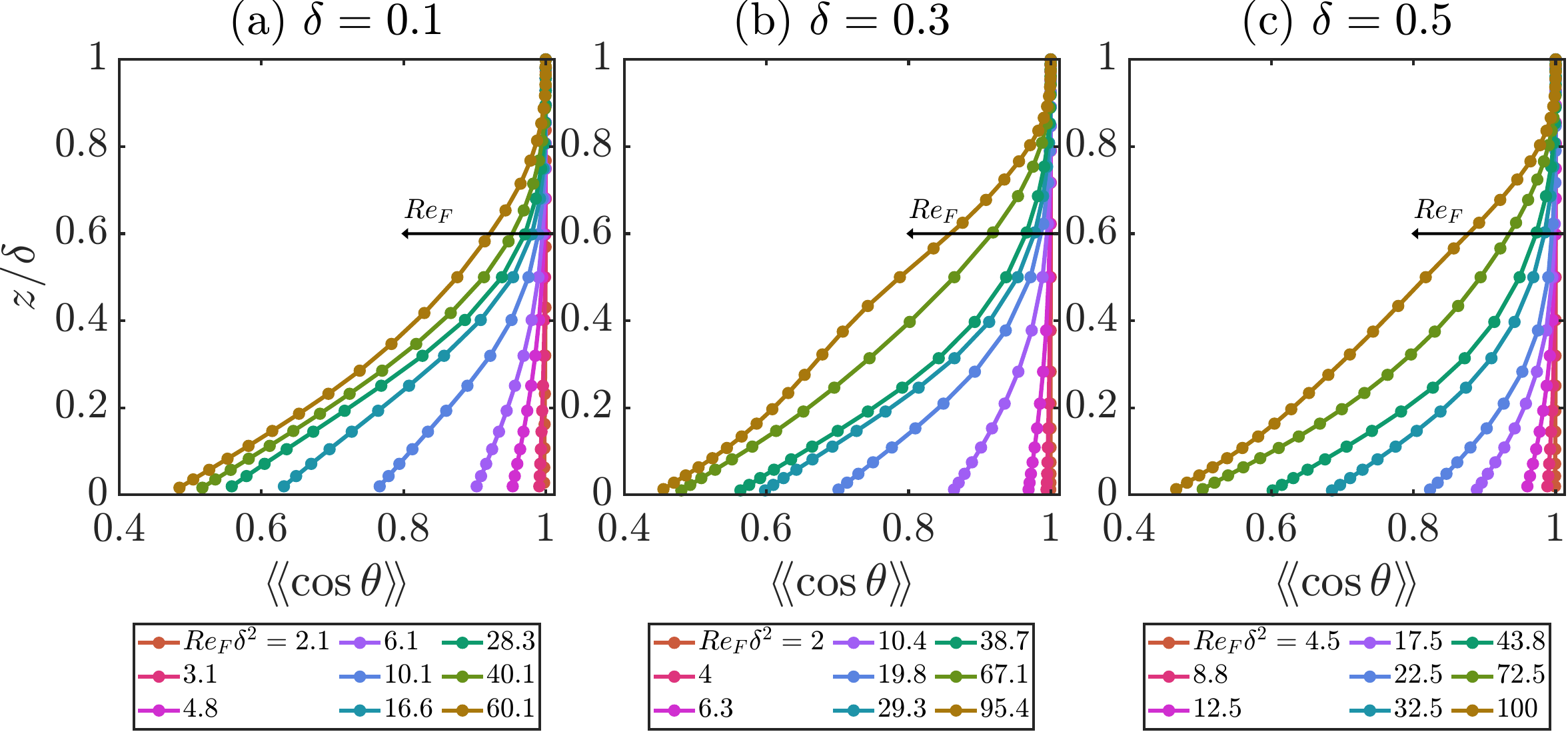}
    \caption{Vertical profiles of $\langle\!\langle \cos\theta \rangle\!\rangle$ for all simulations. Each subfigure corresponds to a different $\delta$ value as indicated by their titles. The arrow in the panels indicates increasing values of $Re_F$.}
    \label{fig:cos0_profiles}
\end{figure}

The behavior of $\cos\theta$ with depth illustrates that flows just below the surface are closely aligned with the surface flow. This alignment remains significant up to a certain depth depending on the value of $Re_F$ (at $\delta$ fixed). This behavior is clearly observed in the vertical profiles of $\langle\!\langle \cos \theta \rangle\!\rangle$ shown in Figure~\ref{fig:cos0_profiles}. For small values of $Re_F\delta^2\lesssim 3$, the subsurface flow remains well aligned with the surface flow throughout the entire layer. For these small values of $Re_F\delta^2$, the vertical velocity profile is Poiseuille-like (and $z_b\approx 0.7\delta$) and the vertical dependence of the horizontal velocity shows a decreasing magnitude, but the orientation of the velocity vector remains almost unchanged. However, as $Re_F\delta^2$ increases, the degree of misalignment intensifies in the lower parts of the fluid layer, leaving only the flow in the upper part with a relatively strong alignment with the surface flow. In addition, the flow gradually exhibits more rapid spatial and temporal variations with increasing $Re_F\delta^2$, which further contributes to the misalignment between surface and subsurface flows.

\subsection{Surface dispersion versus deep dispersion} \label{sec:particle_comparison}
In the previous section, we compared the flow field at several depths in the fluid layer from an Eulerian perspective. Here, we will go one step further by introducing a Lagrangian perspective. We will compare the dispersion characteristics of the passive particles released on the surface with the horizontal dispersion properties of depth-keeping particles embedded in deeper layers of the fluid. We start with a qualitative exploration of horizontal dispersion in Sect.~\ref{Sec4.1}. Subsequently, in Sects.~ \ref{Sec4.2} and \ref{Sec4.3}, we will introduce quantitative measures such as the particle pair correlation function and the correlation dimension to quantify the geometric properties of particle distributions at several heights in the fluid layer, and a measure for the vertical correlation of particle distributions. With these quantitative measures, we will be able to provide a regime diagram in terms of $(z/\delta, Re_F\delta^2)$ and distinguish four different regimes that characterize the Lagrangian dispersion properties of passive particles released at different heights in the fluid layer; see Sect.~\ref{Sec4.5}. In this analysis, we focus exclusively on flows within the inertial regime ($Re_F\delta^2 \gtrsim 10$), characterized by significant temporal fluctuations, specifically when the rms velocity fluctuations are larger than 10\% of the total {\color{black}rms velocity (see Fig. 5 of Ref.~\cite{FloresRamirez2025AsymmetricFlows}).}

\subsubsection{Qualitative comparison of particle distributions}\label{Sec4.1}

\begin{figure}
    \centering
    \includegraphics[scale=0.4]{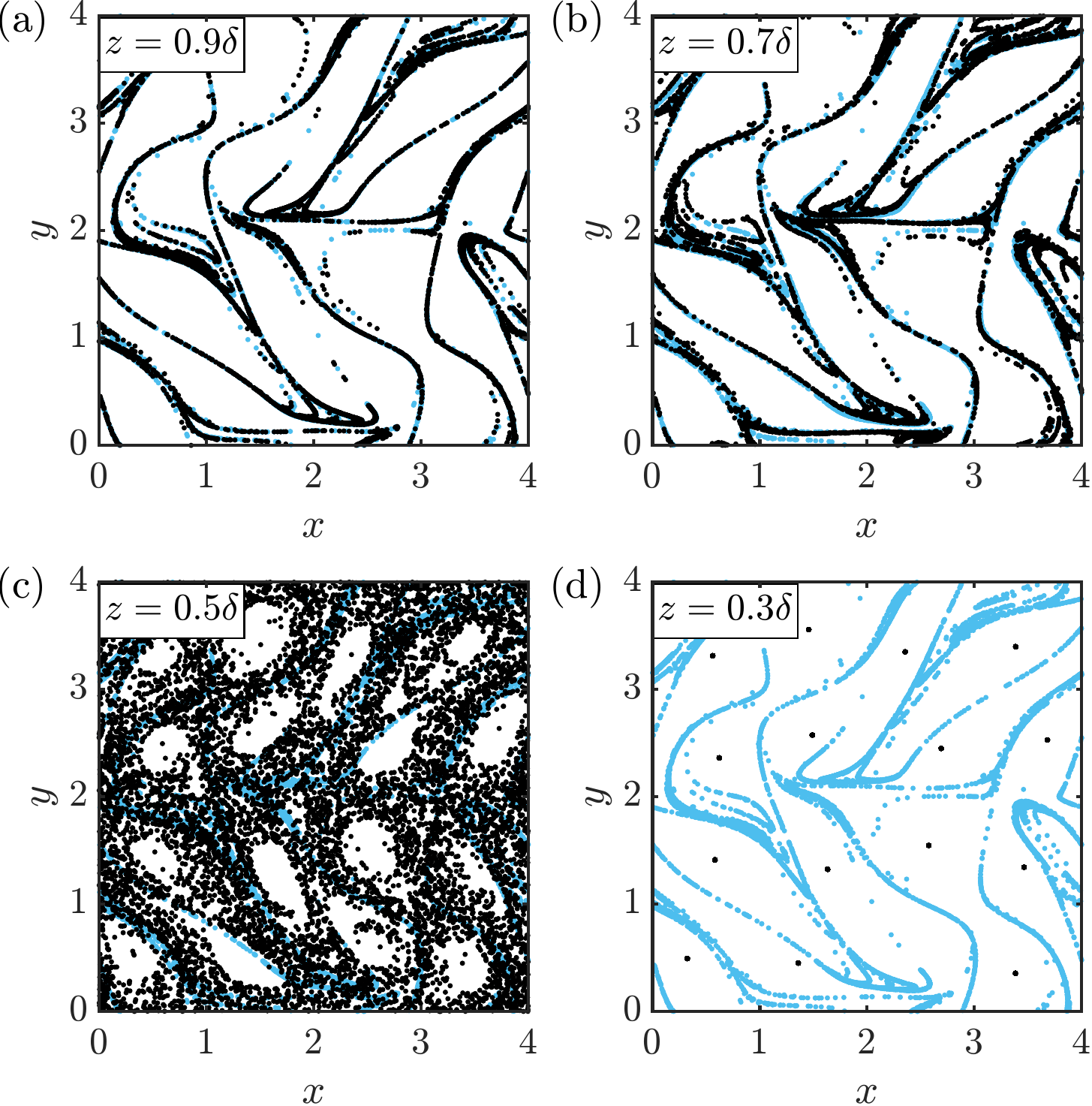}
    \caption{Horizontal distribution of particles (black dots) at selected depths (as shown in the insets) at the end of the simulation. Each panel also shows the distribution of surface particles (blue dots) for comparison. The particles are immersed in a flow with $Re_F=220$ and $\delta=0.3$ ($Re_F\delta^2=19.8$).}
    \label{fig:particle_positions_ReF220_delta03}
\end{figure}

We begin by analyzing the results for particles released at different heights in the fluid layer, with the particles always moving at a fixed depth $z$ (see Sect.~\ref{sec:lagrangian_track}). Figure \ref{fig:particle_positions_ReF220_delta03} shows snapshots of the particle distributions at various depths in the flow with $Re_F=220$ and $\delta=0.3$ ($Re_F\delta^2=19.8$). Particles on the surface (represented by light blue dots) are also included in each panel for comparison. 

The particles, initially distributed uniformly, accumulate in thin, elongated filaments at $z=0.9\delta$ and $0.7\delta$, see Figs.~\ref{fig:particle_positions_ReF220_delta03}a and b. The particle distributions at these depths closely resemble those observed on the surface, especially in terms of the alignment of filaments and location of voids. In contrast, at $z=0.3\delta$ (Fig.~\ref{fig:particle_positions_ReF220_delta03}d), the particles form point-like clusters. However, Fig.~\ref{fig:particle_positions_ReF220_delta03}c shows that at $z=0.5\delta$ the particles remain more or less uniformly distributed, although outside the vortex cores.  

{\color{black}Particle accumulation in filaments or point-like clusters is explained as follows. Particles gather in flow regions where the horizontal divergence is negative, i.e., convergence regions \citep{Cressman2004}. Due to the depth-keeping nature of the particles, they cannot escape from these regions by moving vertically. However, to explain the transition in the geometry of the particle distribution observed in Fig.~\ref{fig:particle_positions_ReF220_delta03}, for a flow with $Re_F \delta ^2=19.8$, one should consider the secondary circulation that develops approximately within each (primary) vortex induced by the forcing. This circulation emerges from the cyclostrophic balance in the presence of bottom drag. In a shallow axisymmetric vortex, the cyclostrophic balance is given by}
{\color{black}
\begin{equation}
    \frac{V^2}{r} = \frac{1}{\rho}\frac{\partial P}{\partial r}~, 
\end{equation}}
{\color{black}with $V$ the swirl velocity of the vortex, $r$ the radial distance from its center, and $\partial P/\partial r$ the radial pressure gradient. The bottom drag weakens the flow near the bottom, whereas the flow remains stronger near the free surface, producing vertical shear in the swirl velocity \citep{Satijn2001,Duran-Matute2010DynamicsVortices,Kamp2012}. Because the cyclostrophic balance must hold at each depth, a faster swirl near the surface requires a stronger radial pressure gradient than for the weaker swirl near the bottom. This depth-dependent radial pressure gradient necessarily implies a vertical pressure gradient, which cannot be balanced by purely swirling motion, and therefore drives a weak secondary circulation. Specifically, the fluid moves radially inward along the bottom, rises near the vortex center, then moves radially outward near the free surface, and finally returns downward along the vortex periphery.} 

{\color{black}In the flow with $Re_F \delta^2=19.8$, the presence of inertial effects introduces fluctuations that prevent the circulation described above from fully being realized; nevertheless, it remains a useful framework for interpreting the transition in the distribution of particles, as shown in Fig.~\ref{fig:particle_positions_ReF220_delta03}. Thus, because radial inflow toward the vortex center occurs close to the bottom, particles collect there, in a manner analogous to the tea leaves paradox \citep{einstein1926}, and eventually collapse into point-like clusters located at the vortex centers. In contrast, particles near the free surface are displaced by the radial outflow toward the vortex edges, forming particle filaments that surround the vortex cores. The coexistence of radial outflow at the surface and radial inflow at the bottom necessarily imply that near mid-depth the radial flow approximately vanishes and hardly any point-like or filamentous particle accumulation will occur.}

\begin{figure}[t!]
    \centering
    \includegraphics[scale=0.4]{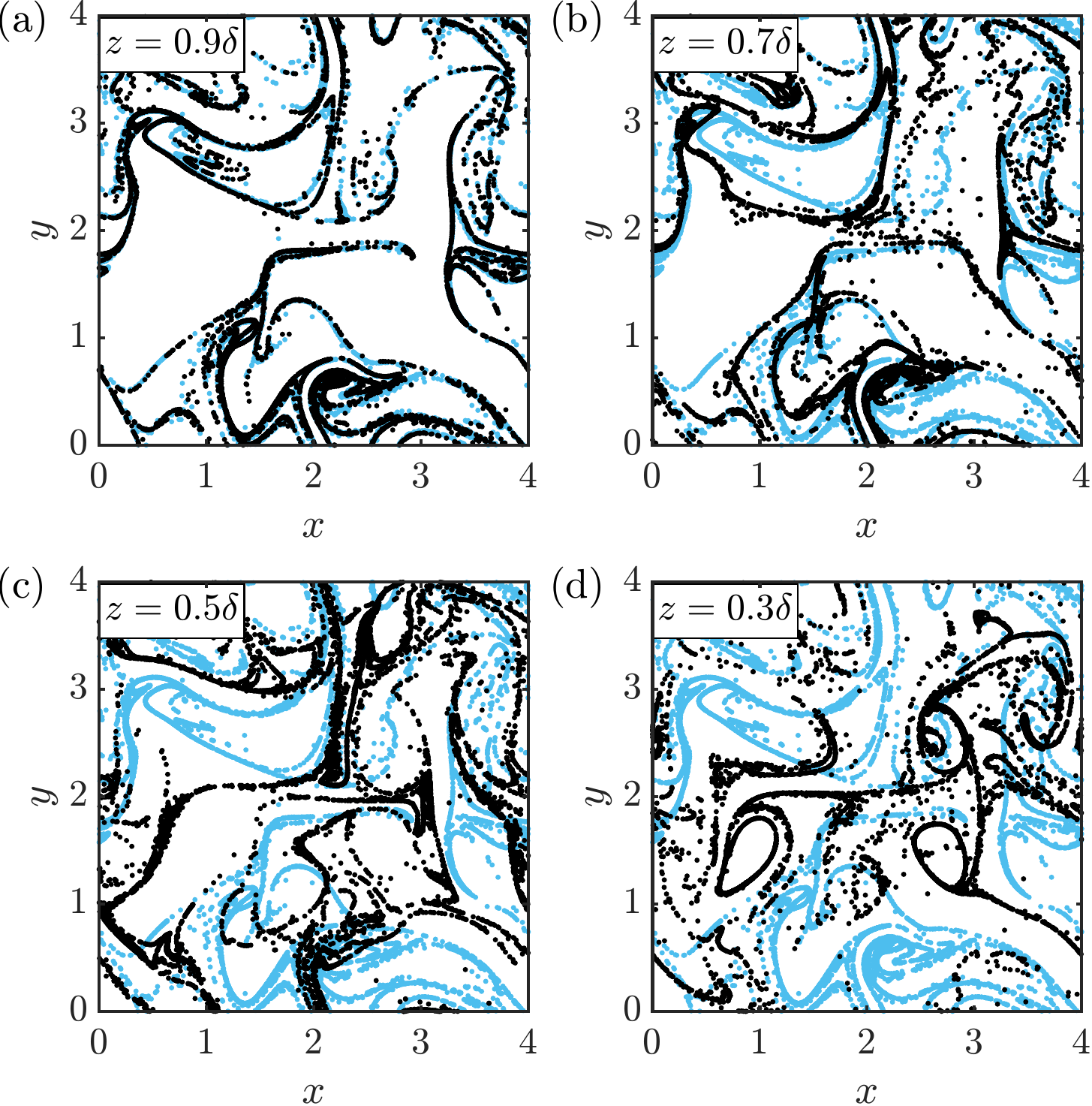}
    \caption{As Fig.~\ref{fig:particle_positions_ReF220_delta03}, except here particles are immersed in a flow with $Re_F=1060$ and $\delta=0.3$ ($Re_F\delta^2=95.4$).}
    \label{fig:particle_positions_ReF1060_delta03}
\end{figure}

To contrast these observations, we also present the particle distributions in an inertia-dominated flow with $Re_F=1060$ and $\delta=0.3$ ($Re_F\delta^2=95.4$), see Fig.~\ref{fig:particle_positions_ReF1060_delta03}. As in the previous case, thin, elongated patches of particles emerge as a result of their preference to accumulate within convergence regions of the flow. In this case, these thin elongated filaments and voids emerge as deep as just above the bottom boundary layer (see the particle distribution at $z=0.3\delta$). Moreover, the particle distribution on the surface and at $z=0.9\delta$ remains similar. However, at larger depth, the geometrical features remain similar, but the location and orientation of the filaments and voids start to decorrelate from the surface pattern.

\subsubsection{Vertical variation of horizontal dispersion}\label{Sec4.2}

To quantitatively characterize the particle distributions described previously, we calculate the correlation dimension, introduced by \citet{Grassberger1983CharacterizationAttractors}, which provides a measure of the geometry of the particle cloud. {\color{black}For this purpose, we compute the correlation function $C(r,t)$ defined as \citep{Grassberger1983CharacterizationAttractors}:}
\begin{equation}
C(r,t)=\frac{1}{M_{\mathrm{pair}}} \sum_{\substack{p,q=1 \\ p\ne q}}^{M_{\mathrm{pair}}} {\mathcal{H}}(r-||{\bf{x}}_h^p(t)-{\bf{x}}_h^q(t)||)~,
\end{equation}
{\color{black}where ${\bf{x}}_h^{p}=(x^{p},y^{p})$ is the horizontal position of particle $p$ (and a similar definition for ${\bf{x}}_h^{q}$), $r$ is a chosen distance, ${\mathcal{H}}$ is the Heaviside step function. Here, $M_{\mathrm{pair}}=M_p (M_p-1)/2$ is the total number of pairs of particles, with $M_p=10^4$ particles randomly selected from the entire set of $N_p$ particles for computational efficiency. In other words, $C(r,t)$ is defined as the (normalized) number of pairs of particles separated by a distance smaller than $r$.} 

\begin{figure}
    \centering
    \includegraphics[scale=0.4]{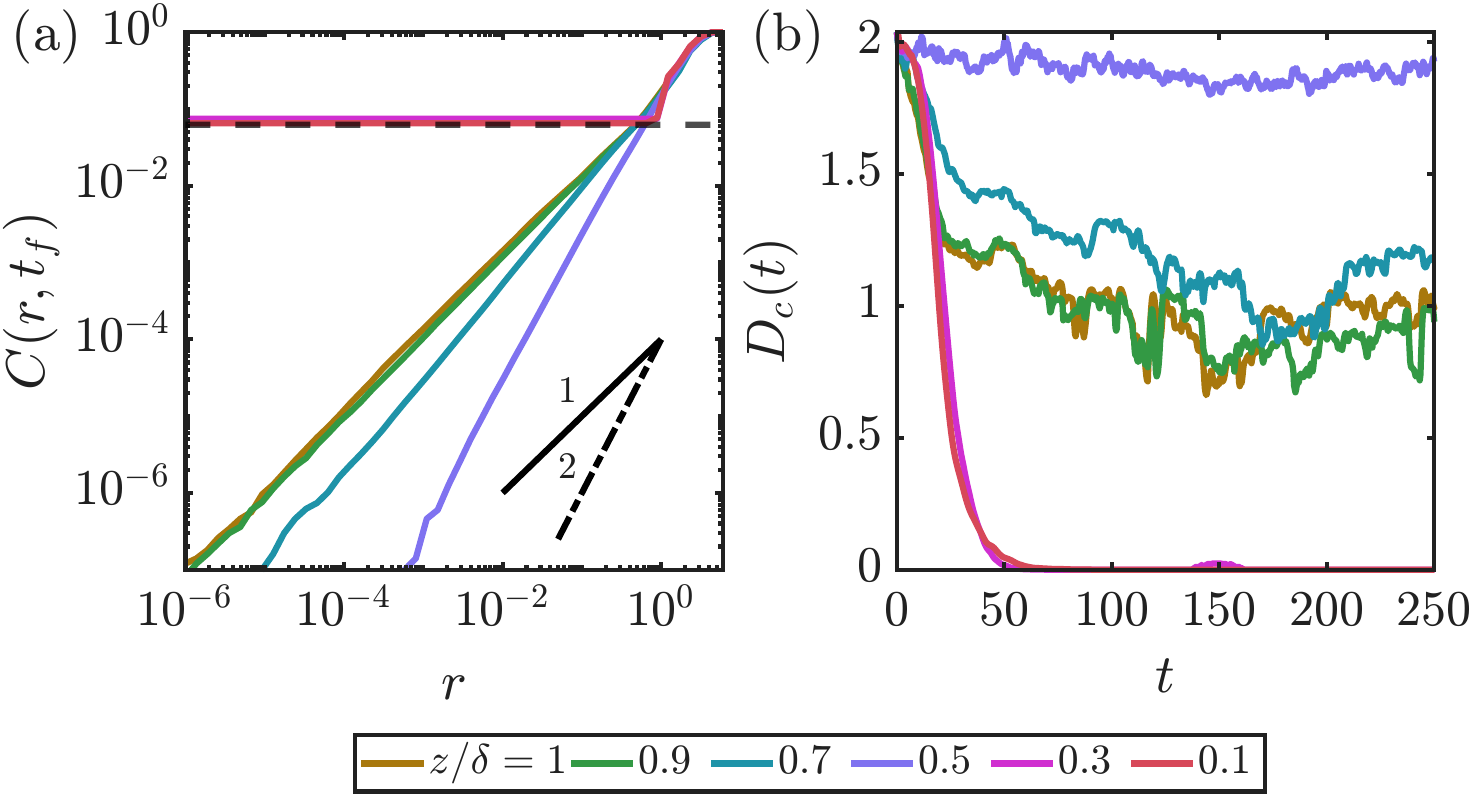}
    \caption{(a) {\color{black}Correlation function evaluated at the final simulation time $t_f$, $C(r,t_f)$, for selected} depths $z/\delta$, derived from the final particle distributions in a flow with $Re_F=220$ and $\delta=0.3$ (see Fig.~\ref{fig:particle_positions_ReF220_delta03}). The black solid and dashed-dotted lines, respectively, correspond $r^1$ and $r^2$ for reference. For the horizontal line segments ($z=0.3\delta$ and $0.1\delta$) for {\color{black}$r\lesssim 1$, $C=Ar^0$ with $A \approx 1/16$ (see horizontal dashed line), indicating} a zero-dimensional particle distribution, that is, particles are concentrated in distinct clusters. (b) Correlation dimension $D_c(t)$ over time at different depths for particles under the same flow conditions.}
    \label{fig:CvsR_DvsTime_ReF220_delta03}
\end{figure}

{\color{black}Figure~\ref{fig:CvsR_DvsTime_ReF220_delta03}a displays the correlation function evaluated at the final simulation time $t_f$, i.e., $C(r,t_f)$, for several depths $z/\delta$. Although $C(r,t)$ changes during the initial evolution of the cloud, it eventually reaches a statistically steady state, and its behavior at $t_f$ is representative of this state. Note that for $r \lesssim 1$, the correlation function curves follow a power-law scaling $C(r,t_f) \propto r^{\alpha}$, with $\alpha$ taking values between one and two (see the auxiliary lines in Fig.~\ref{fig:CvsR_DvsTime_ReF220_delta03}a). An exception is found for $z = 0.1\delta$ and $0.3\delta$, where $C(r,t_f )\propto r^0$. For an explanation, we need to realize that $C(r,t)$ is a cumulative count of pair of particles. For $r \lesssim 10^{-6}$, all counted pairs correspond to particles within the same cluster, since particles inside each of the 16 clusters (see Fig.~\ref{fig:particle_positions_ReF220_delta03}d) are extremely close to each other. For $10^{-6}\lesssim r\lesssim 1$, $C(r,t)$ remains flat because no pair of particles is found at these intermediate distances. Since only intra-cluster pairs have been counted so far, and there are 16 similar clusters, the plateau value is expected to be $C\approx 1/16$ (see the horizontal dashed line in Fig.~\ref{fig:CvsR_DvsTime_ReF220_delta03}a). At $r\approx 1$, i.e., the smallest inter-cluster distance, $C(r,t)$ increases abruptly when inter-cluster pairs begin to contribute and it reaches the value $1$ around $r=3\sqrt{2}$ (largest inter-cluster distance) once all possible pairs have been counted.}

{\color{black}The correlation function $C(r,t)$ exhibits a power-law scaling within a finite range of $r$,
\begin{equation}
    C(r,t)\propto r^{D_c(t)}~, \label{eq:corr_dimension}
\end{equation}
and $C(r,t_f)$ shown in Fig. \ref{fig:CvsR_DvsTime_ReF220_delta03}a represents a special case. According to Grassberger and Procaccia \citep{Grassberger1983CharacterizationAttractors}, the slope $D_c(t)$, which is time dependent as the particle cloud evolves, can be interpreted as the (correlation) dimension that characterizes the geometry of the particle cloud over that range of $r$. In general, $D_c(t)\approx 0$ corresponds to a point-like cluster, $D_c(t)\approx 1$ to a filament-like structure, and $D_c(t)\approx 2$ to an area-filling cloud; fractional values of $D_c(t)$ capture the whole panoply of intermediate structures \citep{Pierrehumbert_1991_GeophysAstrophysFluidDyn}. We obtain $D_c(t)$ by fitting Eq.~(\ref{eq:corr_dimension}) to the computed curves of $C(r,t)$ in the range $10^{-3}\le r\le 10^{-1}$.} 

{\color{black}We consider the correlation dimension to be a more suitable diagnostic for our problem than simpler metrics such as the variance of the particle cloud, which is too coarse to capture differences in the geometry of the particle cloud. For example, a cloud concentrated along the circumference of a disk can exhibit greater variance than a cloud that uniformly fills a disk, although it being less “mixed” \citep{Pierrehumbert1991Large-scaleAtmospheres}. In contrast, the correlation dimension captures this distinction because it reflects how the cloud is spatially organized.}

The temporal evolution of $D_c(t)$ is shown in Fig.~\ref{fig:CvsR_DvsTime_ReF220_delta03}b. For $t\gtrsim 50$, each curve stabilizes around a value between zero and two. {\color{black}For $z = 0.1\delta$ and $0.3\delta$, $D_c(t) \approx 0$ due to the clustering of particles at discrete points (see Fig.~\ref{fig:particle_positions_ReF220_delta03}d). For $z = \delta$, $0.9\delta$ and $0.7\delta$, $D_c (t) \approx 1$ because the particles form well-separated filamentous patches (see Figs.~\ref{fig:particle_positions_ReF220_delta03}a and b). Meanwhile, $D_c (t) \approx 2$ at $z = 0.5\delta$ because the cloud of particles approximately fills the surface (see Fig.~\ref{fig:particle_positions_ReF220_delta03}c).} At $t=0$, $D_c=2$ for all curves because the particles are initially distributed uniformly on a surface. The values obtained for $D_c(t)$ are therefore consistent with the particle patterns observed in Fig.~\ref{fig:particle_positions_ReF220_delta03}.

\begin{figure}
    \centering
    \includegraphics[scale=0.28]{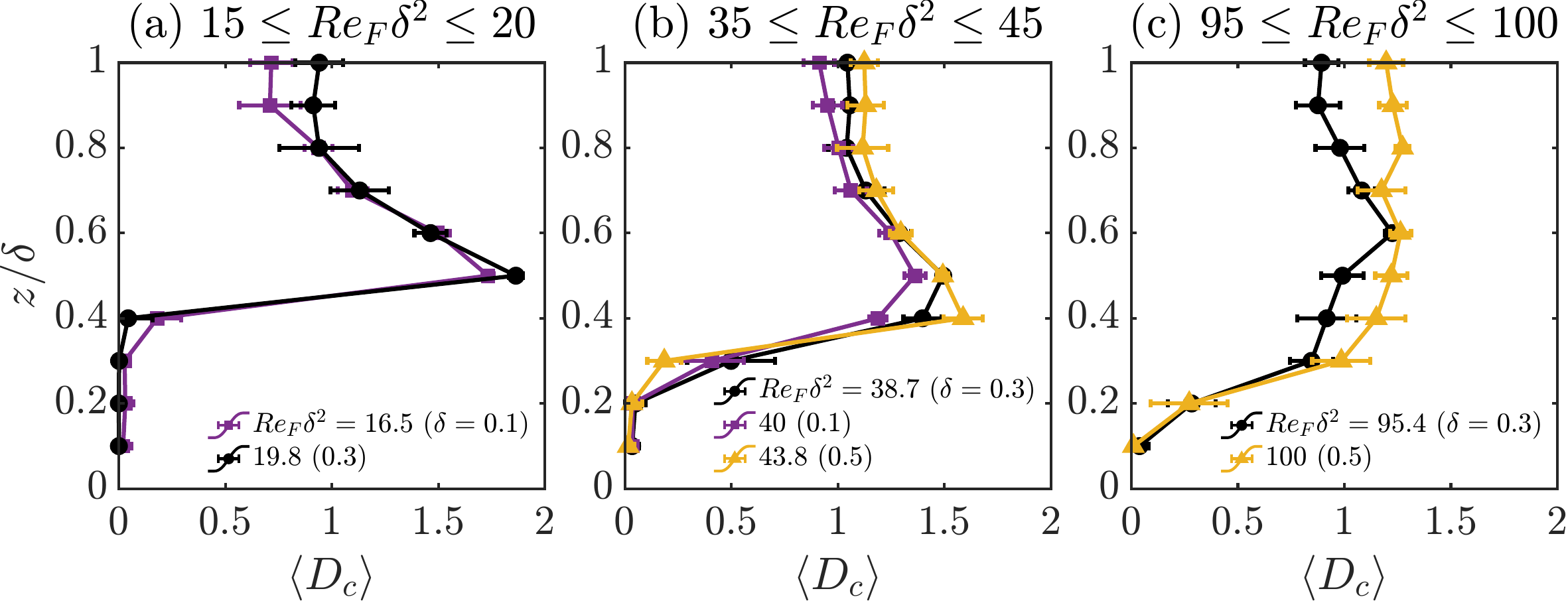}
    \caption{Vertical profiles of the time-averaged correlation dimension $\langle D_c\rangle$ for particles in varying flow conditions characterized by the parameter $Re_F\delta^2$. Each subfigure corresponds to a specific interval of $Re_F\delta^2$, as indicated by their titles.}
    \label{fig:Dc_profiles}
\end{figure}

Since $D_c(t)$ remains approximately steady {\color{black}for $t\gtrsim 100$}, we can obtain a representative value of the correlation dimension by taking its temporal average at a particular height in the fluid layer,
\begin{equation}
    \langle D_c\rangle=\dfrac{1}{\Delta t}\int_{t_0}^{t_0+\Delta t}D_c(t)\,\mathrm{d}t,  \label{eq:time_average2}
\end{equation}
where $t_0$ is an arbitrary time (not the same for every simulation) within the time span where $D_c(t)$ starts to fluctuate around a constant value and $\Delta t$ the time window for averaging. A selection of the resulting vertical profiles of $\langle D_c\rangle$ is shown in Fig.~\ref{fig:Dc_profiles}, grouped into three intervals of increasing $Re_F\delta^2$. In general, the profiles exhibit similar trends within their corresponding $Re_F\delta^2$ ranges.

In Fig.~\ref{fig:Dc_profiles}a, the profiles for $10\leq Re_F\delta^2\leq 20$ are shown. We observe the previously described variation of $\langle D_c\rangle$ in depth: $\langle D_c\rangle\approx 1$ on the surface, a gradual increase to $\langle D_c\rangle\approx 2$ in the upper half of the layer and a sharp drop to zero in the lower half. In the range $35\leq Re_F\delta^2\leq 45$, see Fig.~\ref{fig:Dc_profiles}b, this sharp transition becomes smoother, with profiles now showing $\langle D_c\rangle\approx 1.5$ around $z=0.5\delta$.  For even higher $Re_F\delta^2$ values ($90\leq Re_F\delta^2\leq 100$, shown in Fig.~\ref{fig:Dc_profiles}c), the profiles maintain $\langle D_c\rangle\approx 1$ throughout the layer, particularly for $z \gtrsim 0.3\delta$. This indicates that the particles accumulate predominantly in filamentous patches, an observation supported by Fig.~\ref{fig:particle_positions_ReF1060_delta03}. 

{\color{black} The vertical profiles of $\langle D_c\rangle$ in Fig.~\ref{fig:Dc_profiles} characterize the geometric distribution of the particle clouds as the depth increases. However, they do not quantify the difference between the position of the particle clouds at a certain depth in the fluid layer and those at the surface. For example, when $Re_F \delta^2\approx 100$, $\langle D_c \rangle$ suggests that filamentary structures persist from the surface down to $z \approx 0.3\delta$ (Fig.~\ref{fig:Dc_profiles}c), but it does not reveal whether these deeper filaments resemble those at the surface, which they do not in this case; see Figs.~\ref{fig:particle_positions_ReF1060_delta03}a-c.  Similarly, for $Re_F \delta^2 \approx 20$, filamentary patterns persist down to $z \approx 0.7\delta$ (Fig.~\ref{fig:Dc_profiles}a), yet $\langle D_c \rangle$ alone does not capture the differences between surface and deeper filamentary distributions. Although Fig.~\ref{fig:particle_positions_ReF1060_delta03}c shows a strong qualitative similarity between the two distributions, subtle differences may still exist and therefore require a quantitative evaluation. This motivates the introduction of an additional statistical measure to quantify the similarity (or lack thereof) between particle patterns at the surface and at different depths. This is discussed in the following section.}
 
\subsubsection{Horizontal alignment of particle distributions}\label{Sec4.3}

The vertical correlation between particle distributions in deep layers with those on the surface is explored by analyzing the existence of filaments in deep layers and, if present, their (mis)alignment with those at the surface. For this purpose, we need to introduce a quantitative {\color{black}statistical} measure as a proxy for this vertical correlation. This measure is based on subdividing the horizontal domain into $M^2$ bins with positions $(i,j)$ (with $1\leq i\leq M$ and $1\leq j \leq M$), and counting the number of particles within each bin, denoted as $N_{ij}$. Each bin has a size $\Delta=L/M\approx 0.033$, with $M=120$, and is chosen to be of the same order as the typical size of the mesh elements used in the flow simulations. Note that $\Delta<0.1$ and falls well within the range of length scales used to compute the particle-pair correlation function. From this particle count $N_{ij}$, we define the following indicator function at depth $z$,
\begin{equation}
    I_{ij}(z,t) = \begin{cases}
  1, & \mathrm{if}\; N_{ij}(z,t)\neq 0 ,\\
  0, & \mathrm{if}\; N_{ij}(z,t)= 0,
\end{cases}
\end{equation}
which identifies bins that are occupied or depleted of particles. Both $N_{ij}$ and $I_{ij}$ are computed at each time step for all depths considered. Finally, note that
\begin{equation}
    \dfrac{1}{N_p}\sum_{i,j=1}^MN_{ij}I_{ij}=1.
\end{equation}

\begin{figure}
    \centering
    \includegraphics[scale=0.4]{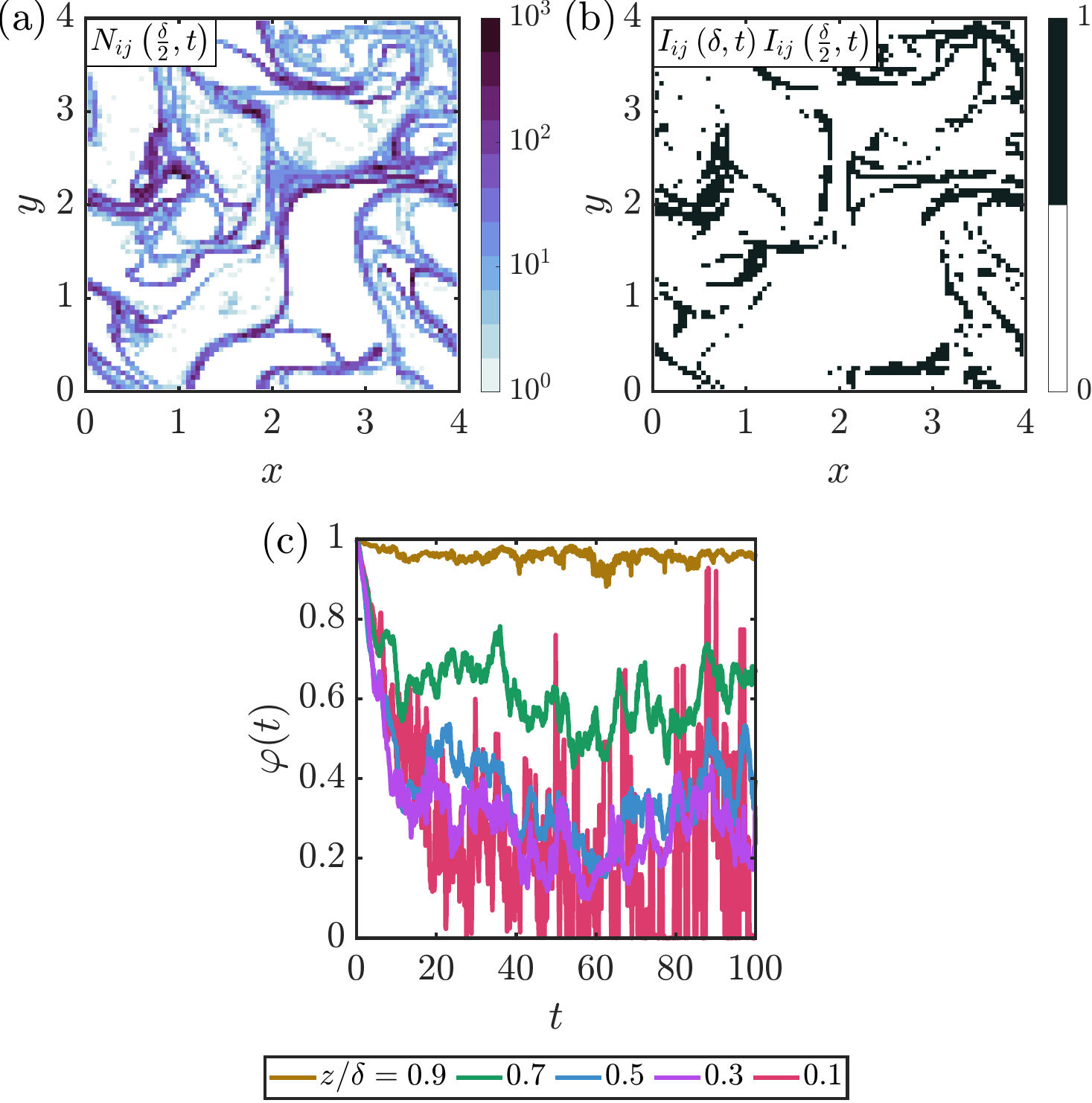}
    \caption{Snapshots of (a) $N_{ij}$ and (b) $I_{ij}(z,t)I_{ij}(\delta,t)$ for $z=0.5\delta$, derived from the final particle distributions in a flow with $Re_F=1060$ and $\delta=0.3$ (see Fig.~\ref{fig:particle_positions_ReF1060_delta03}). (c) Temporal evolution of $\varphi(z,t)$ at different depths for particles under the same flow conditions.}
    \label{fig:DN_DI_PHI_example_ReF1060_D03}
\end{figure}

To quantify the disparity (or similarity) between surface and deep particle distributions, we focus on the spatial alignment of their corresponding occupied and depleted bins. Our approach is based on the fact that, on the free surface, the particle distributions are always of filamentary character, with $\langle D_c\rangle\approx 1$. Those filaments cover a very small part of the free surface and contain almost all particles. This implies that on the free surface, the indicator function $I_{ij}(\delta,t)$ is zero almost everywhere and is only equal to one in bins that overlap with the filaments. We introduce the following index as a proxy for the vertical correlation
\begin{equation}
           \varphi (z,t)=\dfrac{1}{N_p}\sum\limits_{i,j=1}^{M} 
           N_{ij}(z,t)I_{ij}(z,t)I_{ij}(\delta,t). \label{eq:varphi_index}
\end{equation}
Here, the product $I_{ij}(z,t)I_{ij}(\delta,t)$ identifies bins in which particles are present on both the surface and the specified depth simultaneously (see Fig.~\ref{fig:DN_DI_PHI_example_ReF1060_D03}b for a visual example). Thus, $\varphi(z,t)$ measures the fraction of deep particles that horizontally overlap, within the perimeter of the bin, with the surface filaments. It ranges from 0 (no overlap) to 1 (complete overlap). 

{\color{black}The proxy introduced in Eq.~(\ref{eq:varphi_index}) is determined by the observation that the initially homogeneous particle distribution on the free surface (the reference depth for $\varphi$) rapidly evolves toward filamentous clouds of particles. Our proxy would not be suitable when the particles are more or remain uniformly distributed on the free surface, as it could continuously produce misleading values for $\varphi$}. For example, filamentary structures in horizontal planes within the fluid layer will then give {\color{black}$\varphi\approx1$ at all times}. As an alternative, we considered indices based on bin-wise differences in particle counts, but these fluctuate significantly due to the chaotic nature of particle motion, making them unreliable for assessing similarity between distributions.

Figure \ref{fig:DN_DI_PHI_example_ReF1060_D03}c illustrates the temporal evolution of $\varphi(z,t)$ for particles within the flow, with $Re_F=1060$ and $\delta=0.3$ ($Re_F\delta^2=95.4$) at several depths. At $t=0$, $\varphi(z,0) = 1$ because the initial distribution is identical at all depths. Later, $\varphi(z,t)$ decreases and eventually fluctuates around a steady mean {\color{black}value (for $t\gtrsim 20$)}. From Fig.~\ref{fig:DN_DI_PHI_example_ReF1060_D03}c, we see that $\varphi(z=0.9\delta,t)$ remains close to one, indicating nearly identical deep and surface distributions. For lower depths, $\varphi(z,t)$ becomes smaller as more particles at these depths are decorrelated from the surface distribution. For example, at $z=0.5\delta$, only $40\%$ of the deep particles match the surface distribution (since $\varphi(z=0.5\delta,t)\approx0.4$). 

\begin{figure}
    \centering
    \includegraphics[scale=0.28]{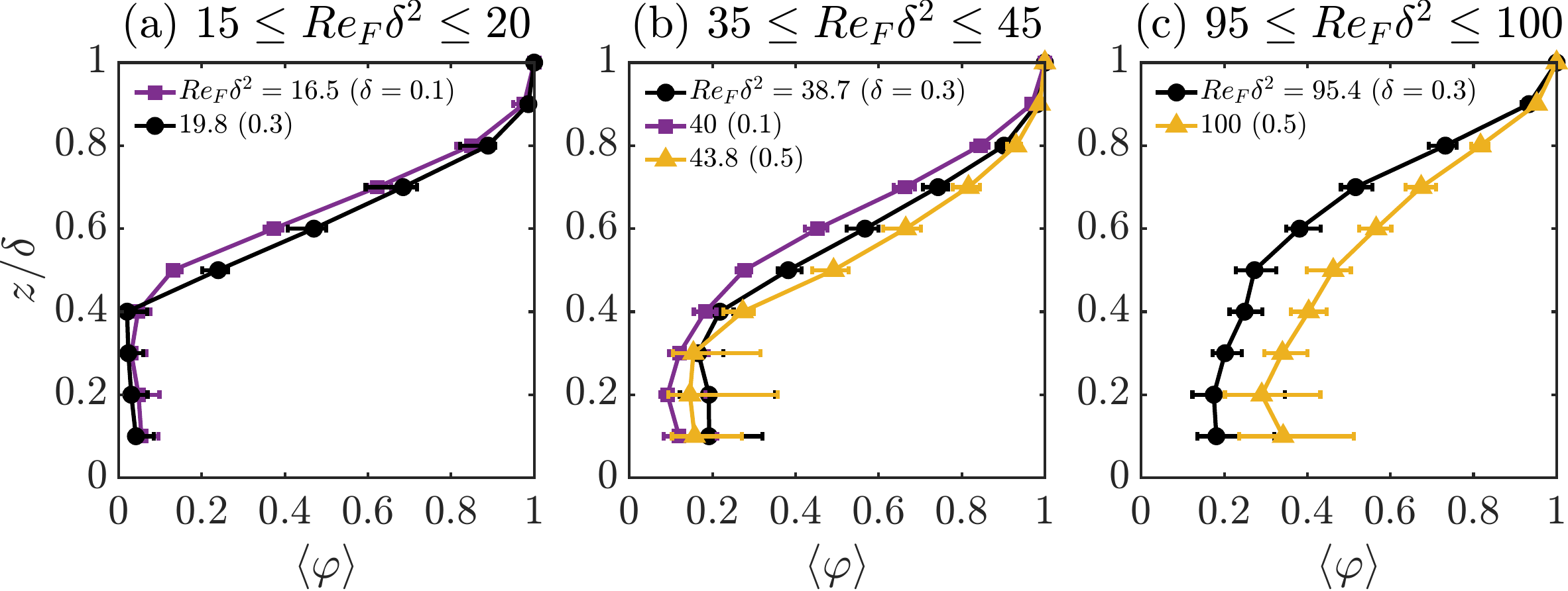}
    \caption{Vertical profiles of $\langle \varphi \rangle$ for particles in different flow conditions characterized by the parameter $Re_F\delta^2$. Each subfigure represents a specific interval of $Re_F\delta^2$, as indicated by their titles.}
    \label{fig:Varphi_profiles}
\end{figure}

Similarly to $D_c(t)$, we construct vertical profiles of $\langle \varphi\rangle$, i.e., a suitable time average of $\varphi(z,t)$ similar to Eq.~\eqref{eq:time_average2}. Figure \ref{fig:Varphi_profiles} shows selected profiles of $\langle \varphi\rangle$, again grouped into three intervals of increasing $Re_F\delta^2$. The profiles exhibit similar shapes within each interval.

In all cases, $\langle \varphi\rangle$ decays monotonically with depth throughout the upper half of the layer, with the decay occurring slightly faster at higher $Re_F\delta^2$. For all values of $Re_F\delta^2$, a portion near the surface ($z \gtrsim 0.8\delta$) exhibits high values, $\langle \varphi\rangle\gtrsim0.8$. In the lower half of the layer, $\langle \varphi\rangle$ is almost independent of depth, with $\langle \varphi\rangle\lesssim0.4$ depending on the range of $Re_F\delta^2$. For smaller $Re_F\delta^2$, we see that $\langle\varphi\rangle\approx 0.01$, reflecting the distinct character of the particle distributions on the surface and near the bottom, where the particles collect in a few spots and $\langle D_c\rangle\approx 0$. For higher $Re_F\delta^2$ values, we still observe filaments and voids above the bottom boundary layer. However, the reduction in $\langle \varphi\rangle$ clearly indicates an increasing misalignment between surface filaments and those at lower depths due to the reorientation of horizontal flows. This is corroborated by the profiles of $\langle\!\langle \cos\theta \rangle\!\rangle$ in Fig.~\ref{fig:cos0_profiles}. Consequently, particles at the surface and in deeper layers are advected in different directions, although at similar speeds (see the profiles of $\langle\!\langle s \rangle\!\rangle$ in Fig.~\ref{fig:sm1_profiles}), promoting spatial misalignment of the particle patches (Fig.~\ref{fig:particle_positions_ReF1060_delta03}c–d).

For lower $Re_F\delta^2$ values ($15\leq Re_F\delta^2 \leq 20$), profiles of $\langle\!\langle \cos\theta \rangle\!\rangle$ in Fig.~\ref{fig:cos0_profiles} show that the horizontal velocities on the surface and at lower depths remain well aligned throughout the layer. Therefore, a similar strong alignment between the particle distributions at different depths might be expected, but this is not observed. The profiles of $\langle\!\langle s \rangle\!\rangle$ in Fig.~\ref{fig:sm1_profiles} exhibit a large vertical shear, which implies large vertical gradients of the components of the horizontal velocity. The impact of these gradients on the particle distributions is evident in Fig.~\ref{fig:Dc_profiles}a (and visually in Fig.~\ref{fig:particle_positions_ReF220_delta03}b–d), where particles are organized into filaments, sheets, and points across a vertical span of roughly $0.4\delta$.  

\begin{figure}
    \centering
    \includegraphics[scale=0.4]{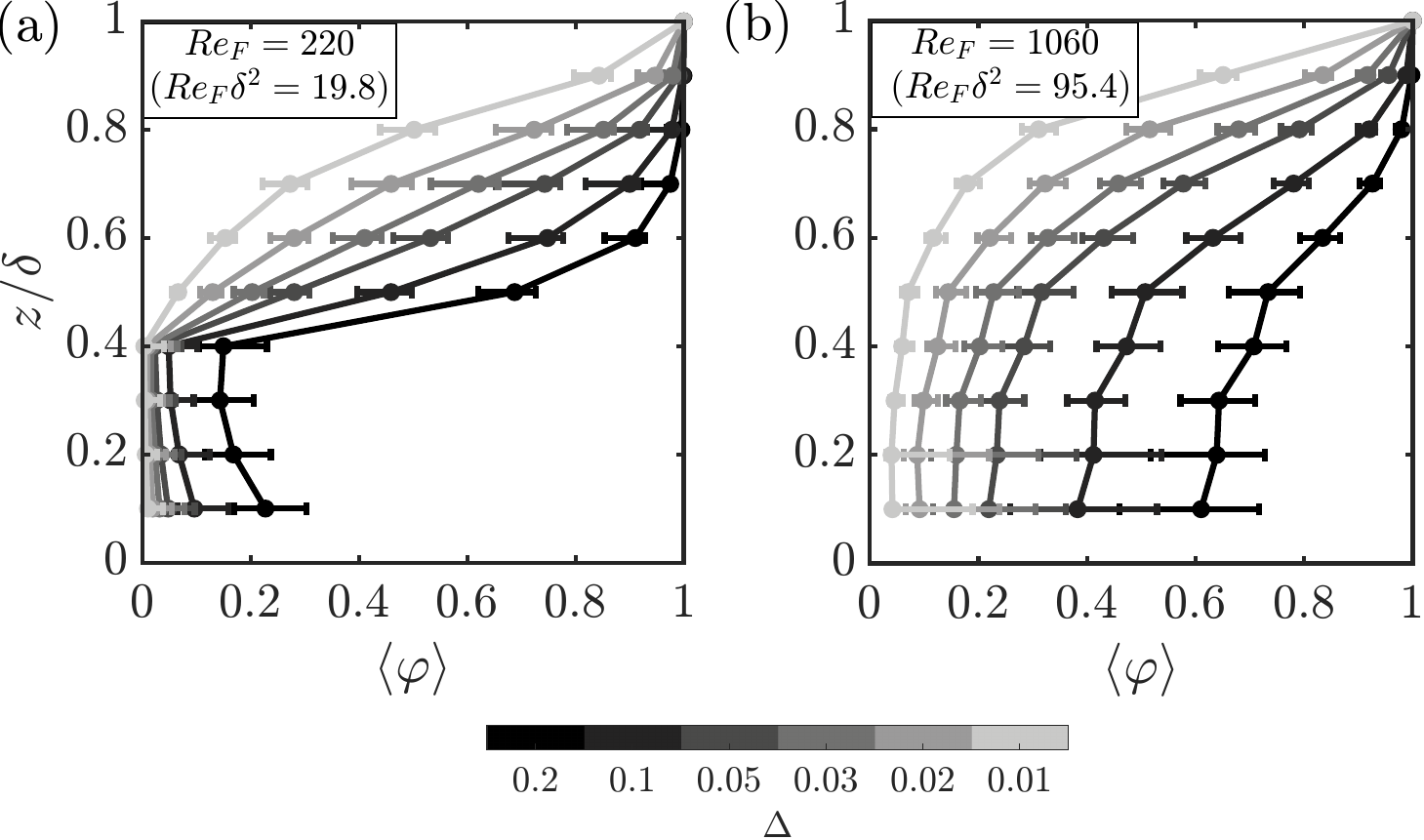}
    \caption{Vertical profiles of $\langle \varphi \rangle$ computed using various bin sizes $\Delta$. The data are derived from particle distributions in flows with $\delta=0.3$ and (a) $Re_F=220$ and (b) $Re_F=1060$.}
    \label{fig:DeltaSens_ReF220_delta03}
\end{figure}

To better interpret $\varphi$ and its average $\langle \varphi\rangle$, we must consider its dependence on the bin size $\Delta$. Figure \ref{fig:DeltaSens_ReF220_delta03} shows $\langle \varphi\rangle$ computed using various $\Delta$ values for the case with $Re_F = 220$ and $\delta = 0.3$ ($Re_F\delta^2=19.8$). Decreasing $\Delta$ systematically reduces $\langle \varphi\rangle$. 
For $\Delta \leq 0.05$, the resolution gradually becomes sufficient to resolve misalignments between the particle distributions for $Re_F=220$. For $Re_F=1060$ it is not yet optimal.

\subsubsection{Impact of vertical motion and turbulence}\label{Sec4.4}

The current investigation focused on the transport of Lagrangian particles constrained to a fixed depth in the fluid column and without the impact of background turbulence. {\color{black}In this section, with the aim to confirm the robustness of the results discussed in Sections \ref{Sec4.2} and \ref{Sec4.3}, we briefly} discuss the consequences of allowing particles to move vertically but still constrained vertically by a restoring force, and the presence of a background turbulent velocity field on horizontal particle transport. We explore both aspects and its impact {\color{black}on such transport processes} by introducing models to represent these phenomena.

We first consider the relaxation of the vertical confinement of the particles. To this end, we compute particle trajectories using a modified version of Eq.~(\ref{eq:eq_motion_particles}),
\begin{equation}
\dfrac{d\textbf{x}^p(t)} {dt}=\mathbf{u}^p_f+w^ p_r \mathbf{e}_z~, \label{eq:eq_motion_particles_vert} 
\end{equation}
where the particles are advected horizontally and vertically by the flow (in contrast to Eq.~(\ref{eq:eq_motion_particles}), $w_f^p\ne 0$), and $w^ p_r \mathbf{e}_z$ represents a vertical restoring velocity, with
\begin{equation}
    w^ p_r(z^p)=-\mathcal{K}(z^p-z_0)  \label{eq:rest_vel}
\end{equation}
and $\mathbf{e}_z$ the unit vector in the vertical direction \cite{DePietro2015ClusteringTurbulence}. In general, this velocity acts to vertically confine particles near a reference depth $z_0$, in this case the release depth. This behavior is analogous to that of zooplanktonic microorganisms or miniature drifters that actively regulate their vertical position {\color{black}through swimming and buoyancy \cite{Franks1992SinkFronts,FloresRamirez2025VerticalFlows,Genin2005SwimmingAggregation,Chen2021DielPatchiness,Jaffe2017ADynamics,Morgan2021}.} The strength of this confinement is determined by the constant $\mathcal{K}$. When $\mathcal{K}\rightarrow\infty$, the particles are perfectly constrained to the plane at $z_0$ where the particles were released, while for $\mathcal{K}=0$, the particles can freely advect throughout the fluid column. The restoring velocity $w^ p_r(z^p)$ experienced by the particle decreases as its vertical position $z^p$ approaches $z_0$. {\color{black}Eqs.~(\ref{eq:eq_motion_particles_vert}) and (\ref{eq:rest_vel}) are formally derived from the Maxey–Riley equations (see \citet{DePietro2015ClusteringTurbulence}) considering almost neutrally buoyant particles in a weakly linearly stratified fluid. In this derivation, Eq. (\ref{eq:rest_vel}) originates from the buoyancy term and reflects the force that would restore the particle to its reference depth (hence the term “restoring”).}  

\begin{figure}
    \centering
    \includegraphics[scale=0.28]{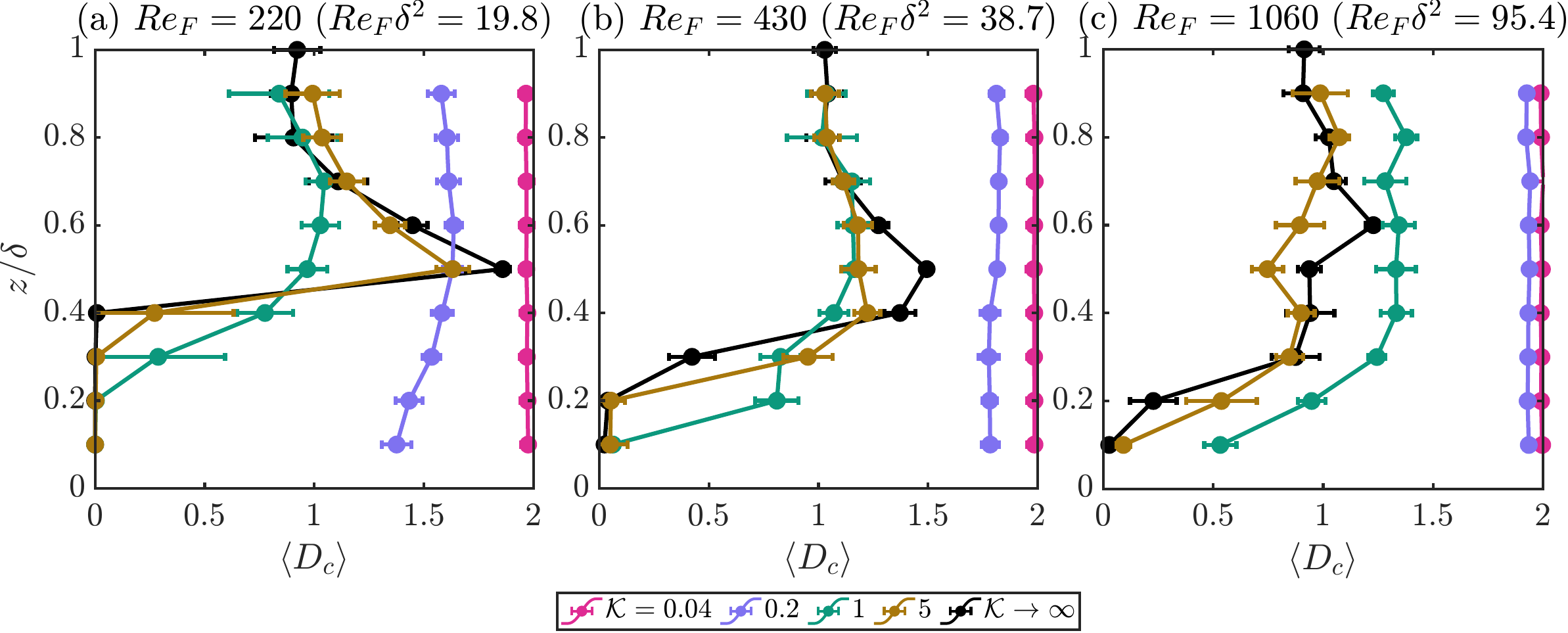}
    \caption{Vertical profiles of the time-averaged correlation dimension $\langle D_c\rangle$ for different values of the constant ${\mathcal{K}}$. The profiles for $\mathcal{K}=\infty$ correspond to those plotted in Fig.~\ref{fig:Dc_profiles}. The particles are immersed in flows with $\delta=0.3$ in varying flow conditions characterized by the parameter $Re_F\delta^2$, as indicated by their titles.}
    \label{fig:Dc_profiles_vert}
\end{figure}

The particles within the fluid column are advected by the full 3D flow and their vertical motion is damped by the restoring velocity. We consider $\mathcal{K}=0.04,0.2,1$ and $5$. In Sect.~\ref{Sec4.2}, we explore how surface dispersion correlates with deeper dispersion; see Fig.~\ref{fig:Dc_profiles} and \ref{fig:Varphi_profiles}, and here we evaluate how the vertical excursion of the particles modulates this correlation. In Fig.~\ref{fig:Dc_profiles_vert} we show the vertical profiles of $\langle D_c\rangle$ for some cases ($Re_F=220$, $430$ and $1060$) of particles immersed in flows with $\delta=0.3$. 

For ${\mathcal{K}}\gtrsim 5$, the time-averaged correlation dimension $\langle D_c\rangle$ resembles that of particles perfectly constrained to the depth of release. For ${\mathcal{K}}=5$ the width of the particle layer, based on the standard deviation of the vertical particle position with respect to the mean position (for this case, close to the depth of release), varies between $0.04\delta$ (for $Re_F=220$) and $0.1\delta$ ($Re_F=1060$). The vertical profiles of $\langle \varphi\rangle$ (not shown here) for ${\mathcal{K}}=5$ are similar to those for ${\mathcal{K}}=\infty$. The cases with ${\mathcal{K}}\lesssim 1$ show that vertical excursions of the particles contribute to an almost uniform horizontal distribution of the particles in the plane of release and $\langle D_c\rangle\rightarrow 2$. The particle distribution is then also decorrelated from the surface distribution, which is supported by $\langle \varphi\rangle\ll 1$. For ${\mathcal{K}}=0.2$, the width of the particle layer varies between $0.32\delta$ (for $Re_F=220$) and $0.44\delta$ ($Re_F=1060$), covering almost half of the fluid column.   

\begin{figure}
    \centering
    \includegraphics[scale=0.28]{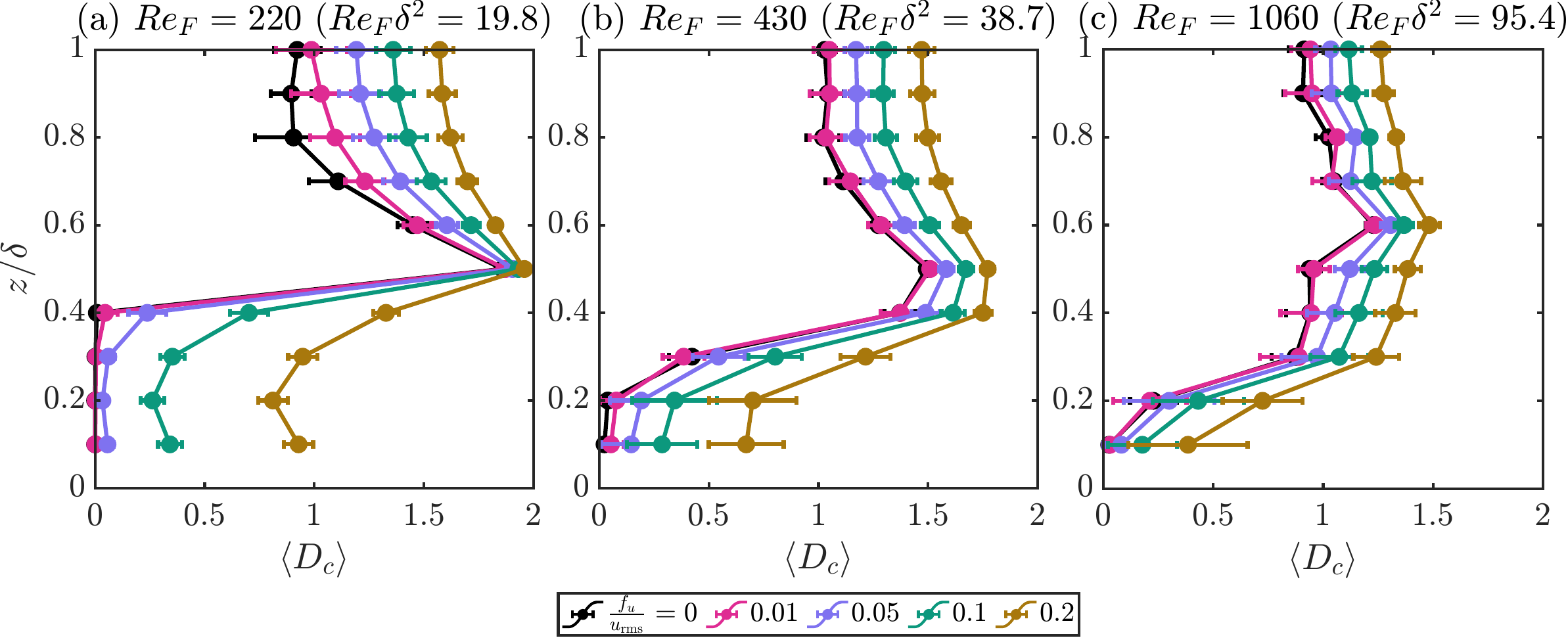}
    \caption{Vertical profiles of the time-averaged correlation dimension $\langle D_c\rangle$ for different values of $f_u$ (parameter that controls the intensity of the turbulent motion, and proportional to $u_\mathrm{rms}$). The particles are immersed in flows with $\delta=0.3$ in varying flow conditions characterized by the parameter $Re_F\delta^2$, as indicated by their titles.}
    \label{fig:Dc_profiles_turb}
\end{figure}

{\color{black}Additionally, we perform Lagrangian} tracking that accounts for the effects of small-scale turbulence on particle motion. This is done by adding a random velocity $\mathbf{u}^p_r=(u_r^p(t),v_r^p(t),0)$, which parameterizes the action of turbulent (incoherent) motions in the horizontal plane, to the (deterministic) velocity of the particles \cite{Zambianchi1994EffectsEstimates}. Thus, the modified equation of particle motion is 
\begin{eqnarray}
     \dfrac{d{\textbf{x}}^p(t)} {dt}={\mathbf{u}}^p_f+{\mathbf{u}}^p_r~, \label{eq:eq_motion_particles_random} 
\end{eqnarray}
considering horizontal motion only. This implies that at each integration step the particle receives a random ``kick'' due to the action of turbulent motions, and each component of ${\mathbf{u}}^p_r$ is drawn from a uniform distribution over the interval $[-f_u,f_u]$, resulting in zero mean and variance $f_u^2/3$ \cite{OcampoJaimes2022DispersionModel}. The parameter $f_u$ controls the intensity of the turbulence and is set as $f_u/u_{\mathrm{rms}}=0.01,0.05,0.1$ and $0.2$, where $u_\mathrm{rms}=\langle\!\langle u^2 \rangle\!\rangle^{1/2}$ represents the rms velocity of the deterministic velocity field on the surface (based on an average over space and time), and the turbulence represents a relatively small disturbance. {\color{black}The random “kicks” used here provide a minimal and idealized description of turbulent fluctuations. This approach assumes a scale separation between the deterministic (numerically resolved large-scale) and stochastic (unresolved small-scale) components of the velocity field. The latter component is treated as turbulent and its contribution to the motion of the particles is thus modeled as a stochastic (random) process uncorrelated in time, similar to the process proposed by \citet{Zambianchi1994EffectsEstimates} and \citet{Griffa1996}.} In Fig.~\ref{fig:Dc_profiles_turb} we show the vertical profiles of $\langle D_c\rangle$ for a few cases ($Re_F=220$, $430$ and $1060$) of particles immersed in flows with $\delta=0.3$ and different intensities of turbulence.

The presence of small-scale turbulence will affect $\langle D_c\rangle$, but its impact is relatively small for $f_u\lesssim 0.05u_\mathrm{rms}$. The conclusions drawn for $\langle D_c\rangle$ and $\langle\varphi\rangle$ are similar to those for $f_u=0$. In all cases, there is a tendency to larger $\langle D_c\rangle$, as can be expected, since turbulent diffusion tends to homogenize the particle distribution in the horizontal plane. This effect becomes clearly visible for $f_u\gtrsim 0.1u_\mathrm{rms}$.

\subsubsection{Four regimes for horizontal particle transport} \label{Sec4.5}

\begin{figure}[t!]
    \centering
    \includegraphics[scale=0.99]{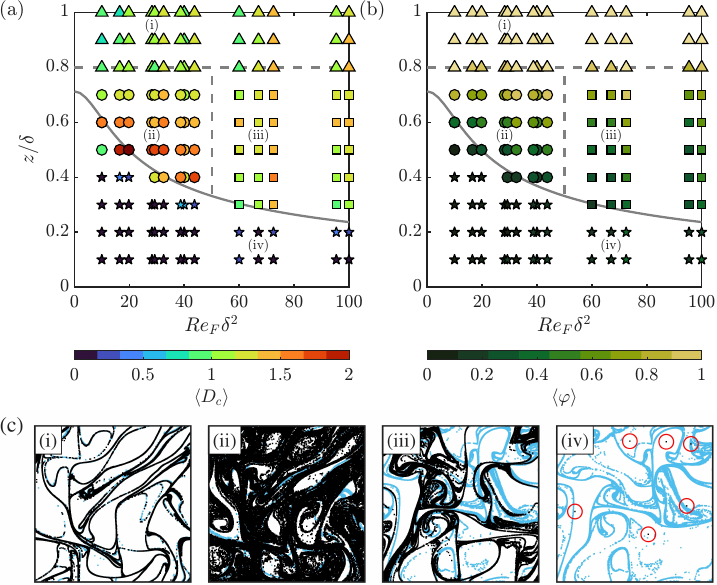}
    \caption{The location of all our inertial regime simulations within the ($z/\delta$,$Re_F\delta^2$) parameter space. The simulations are colored based on (a) $\langle D_c\rangle$ and (b) $\langle \varphi\rangle$  values. The solid line corresponds to the continuous boundary layer thickness given by Eq.~(\ref{eq:zbd}), while the dotted lines correspond to  empirically determined thresholds in $\langle D_c\rangle$ and $\langle \varphi\rangle$. These lines divide the parameter space into four regimes: I (triangles), II (circles), III (squares), and IV (stars). {\color{black}(c) Illustrative cases for each regime, denoted by (i)-(iv), with the corresponding snapshots of particle positions at the surface (blue dots) and at depth (black dots). The red circles in panel (c,iv) indicate the locations of point-like particle clusters.}}
    \label{fig:parameter_space_z_vs_ReFD2}
\end{figure}

We summarize the relation between vertical correlation of particle distributions, estimated by $\langle \varphi \rangle$, and the geometric characteristics of the {\color{black}particle clouds}, quantified by the correlation dimension $\langle D_c\rangle$, with a regime diagram established by $z/\delta$ and $Re_F\delta^2$. {\color{black}This allows us to empirically identify the depth range over which the surface particle distributions remain statistically representative of those at depth, in terms of their geometrical and spatial characteristics.} We plot both $\langle D_c\rangle$ and $\langle \varphi \rangle$ in this regime diagram in Fig.~\ref{fig:parameter_space_z_vs_ReFD2}. We only consider simulations within the inertial regime. Data points of $\langle D_c\rangle$ and $\langle \varphi \rangle$ are colored according to their corresponding values. Based on these values and the behavior of the boundary layer (solid lines in each panel, see also Fig.~\ref{fig:zb_vs_ReFD2}), we can divide the parameter space into four distinct regions. {\color{black}We have included snapshots of particle positions on the surface
(blue dots) and at depth (black dots) in Fig.~\ref{fig:parameter_space_z_vs_ReFD2}c to illustrate the particle distributions in each regime (denoted by the symbols (i)-(iv) in Figs.~\ref{fig:parameter_space_z_vs_ReFD2}a and~\ref{fig:parameter_space_z_vs_ReFD2}b, representing each of these four regimes).}

Regime I consists of particles that form thin, elongated filaments ($0.75\lesssim \langle D_c\rangle\lesssim 1.25$, {\color{black}see Figs.~\ref{fig:particle_positions_ReF220_delta03}a, \ref{fig:particle_positions_ReF1060_delta03}a and Fig.~\ref{fig:parameter_space_z_vs_ReFD2}c,(i)).} Since surface particles also form filaments, the distributions at these depths ($z \gtrsim 0.8\delta$) remain well aligned with the surface distribution because $\langle \varphi \rangle \gtrsim 0.8$. This region spans the entire range of values $Re_F\delta^2$. 

Regimes II and III lie approximately within $z_b\lesssim z\lesssim 0.8\delta$, where the deep particle distributions lose alignment with the surface distributions ($0.3\lesssim \langle \varphi \rangle \lesssim 0.8$). The reasons for this loss differ between these regions, which justifies their distinction. Regime II contains particles that form wide filaments to sheets ($1.5\lesssim\langle D_c\rangle\lesssim2$, {\color{black}see Fig.~\ref{fig:particle_positions_ReF220_delta03}c and Fig.~\ref{fig:parameter_space_z_vs_ReFD2}c,(ii)).} Because these structures differ from surface filaments, the alignment weakens with depth. This behavior is observed in flows with $Re_F\delta^2\lesssim 50$. Regime III features particle filaments similar to those at the surface ($0.75\lesssim \langle D_c\rangle\lesssim 1.25$, {\color{black}see Fig.~\ref{fig:particle_positions_ReF1060_delta03}c and Fig.~\ref{fig:parameter_space_z_vs_ReFD2}c,(iii)).}  However, strong flow fluctuations misalign these deep filaments relative to surface filaments, reducing their alignment. This occurs in flows with $Re_F\delta^2\gtrsim50$, where strong flow unsteadiness becomes dominant.

Finally, regime IV consists mainly of particles organized in isolated points ($0\lesssim \langle D_c\rangle\lesssim 0.3$, {\color{black}see Fig.~\ref{fig:particle_positions_ReF220_delta03}d and Fig.~\ref{fig:parameter_space_z_vs_ReFD2}c,(iv)), that form at depths} within the boundary layer ($z\lesssim z_b$). Because these structures do not resemble surface filaments, deep particle distributions are completely decorrelated from surface distributions ($\langle \varphi \rangle \lesssim 0.3$). Regime IV is clearly separated from regimes II and III and $z_b$, defined in Eq.~(\ref{eq:zbd}), represents the boundary between these regimes and coincides with the large gradients in $\langle D_c \rangle$ and $\langle\varphi\rangle$.

\newpage
\section{Summary and conclusion} \label{sec:conclusion}

{\color{black}This study investigates whether the dispersion of passive particles at the free surface can serve as a reliable statistical proxy for the dispersive behavior of depth-keeping particles within a shallow fluid layer. To this end, we statistically compared the spatial distributions of particles confined at various depths with those at the surface.} The comparison was carried out under varying flow conditions, characterized by the parameter $Re_F\delta^2$, where $Re_F$ is the Reynolds numbers and $\delta$ is the thickness of the fluid layer.

We analyzed the horizontal velocity fields at different depths, focusing on the alignment between surface and deep velocities, and their speed ratio to assess how well surface flows represent deeper ones. The relation between surface and deeper flows was found to depend on the parameter $Re_F\delta^2$. Specifically, at low $Re_F\delta^2$ (viscous regime), the velocity alignment remains strong throughout the entire layer. However, the speed ratio reveals a Poiseuille-like vertical dependence of the flow magnitude. Therefore, deep flows closely follow the surface flow, differing only in magnitude. In contrast, at large $Re_F\delta^2$ (inertial regime), the alignment is well maintained just below the surface, but deteriorates at greater depths. The speed ratio remains nearly constant in the inviscid interior, but decreases within the boundary layer. Then, deep flows match the surface flow in magnitude (above the boundary layer) but not always in direction. Strong flow unsteadiness in this regime leads to decorrelation between surface and deeper flows.

Exploration of the full parameter regime ($z/\delta,Re_F\delta^2$) resulted in the identification of four regimes. These regimes are summarized below.

Regime I: In the upper quarter of the fluid layer, the horizontal particle distributions correlate strongly with those on the surface, regardless of {\color{black}the $Re_F\delta^2$ values considered.} To support this conclusion, we employ two key {\color{black}statistical measures:} $\varphi$, which quantifies the degree of alignment between the positions of surface and deep particles, and the correlation dimension $D_c$, which characterizes the geometry of the horizontal particle distributions. Within this depth range, the observed correlation arises from both surface and deep particles forming thin, elongated filamentary structures whose locations are closely aligned.

Below the upper quarter of the fluid layer, the correlation between horizontal particle distributions in the deeper layers with those on the surface progressively weakens. The underlying mechanisms driving this decay vary with $Re_F\delta^2$.

{\color{black}Regime II: For moderate values ($10 \lesssim Re_F\delta^2 \lesssim 50$), the correlation is reduced primarily due to the transition in the particle patches from filamentary to more diffuse or even quasi-uniform distributions at mid-depth. This transition is a result of the secondary circulation that takes place in the vortices: radial inflow near the bottom drives particle clustering at the vortex centers, whereas radial outflow near the surface advects particles toward the vortex periphery, forming filaments. The almost absence of these opposing radial flows near mid-depth implies that the radial velocity approximately vanishes there, thus suppressing both point-like particle clustering and filament formation, and classical horizontal dispersion is recovered.}

Regime III: In contrast, at $Re_F\delta^2\gtrsim50$, filamentary structures persist at greater depths, resembling those of the surface. However, because of pronounced flow unsteadiness, these structures become spatially mismatched, reducing the overall correlation between surface and deep particle distributions. 

Regime IV: Within the bottom boundary layer, deep particles cluster into isolated points, which causes their distribution to become fully uncorrelated with the surface distribution.

In summary, in regime I, we can infer subsurface transport processes quantitatively from surface observations. In regime III, vertical correlation is gradually lost, but subsurface advective horizontal transport of particles is qualitatively similar to transport on the surface.

We have considered an idealized situation of passive particles embedded in a shallow unsteady laminar flow field without the presence of small-scale turbulence. Additionally, the vertical confinement of particles to a specific horizontal plane is usually very constraining. {\color{black}As a first preliminary exploration to assess the robustness of our results in Sections \ref{Sec4.2} and \ref{Sec4.3}, and associated regime diagram,} we have included a proxy for horizontal turbulent diffusion and allowed limited vertical excursions of the particles to explore the impact on (horizontal) Lagrangian transport. {\color{black}Another possibility, but not pursued in the present study, is to add more features to the model used to constrain the vertical excursion of the particles (i.e., the depth-directed swimming). For example, one could incorporate body rotations induced by local velocity gradients, allowing swimmers to orient their swimming direction in response to flow features (see, for instance, \citet{DeLillo2014} and \citet{Ventrella2023}). In addition, alternative parameterizations of turbulent fluctuations acting on the particles could be investigated; for example, in contrast to the approach used here, the turbulent contribution to particle motion could be modeled as a temporally correlated stochastic process, and progressively extended toward more complex formulations; see \citet{Zambianchi1994EffectsEstimates} and \citet{Griffa1996}.} 

{\color{black}Our preliminary exploration of horizontal stochastic impulses (to mimic the effects of turbulence) or allowing vertical motion with a restoring force shows that the main conclusions} remain valid (for the set of simulations considered) when the intensity of the turbulence is relatively low and when the particles are constrained to thin slabs around the depth of release. However, a deeper exploration is warranted. Open questions include competition between horizontal randomization of particles by turbulence and the flow convergence in the horizontal plane leading to the formation of filamentary particle patches, and the impact of relaxing the vertical confinement of the particles on this formation process. {\color{black}Moreover, the forcing used in our simulations is purely horizontal and time-independent, unlike electromagnetic forcing in laboratory experiments and real environmental forcing such as due to wind stress. Our prime concern is the impact of quasi-2D flows (with weak secondary circulations) or 3D laminar flows (with significantly stronger secondary circulations) on Lagrangian horizontal transport of particles at different heights in the shallow fluid layer. The forcing is a means of generating the quasi-2D or 3D laminar flow, and different forcing schemes under the present parameter settings ($Re_F\delta^2$) will still generate quasi-2D or 3D laminar flows. However, further exploration with more complex, time-dependent, and three-dimensional forcing is an interesting next step. In environmental flows, we cannot define $Re_F$, but using $U_{\mathrm{rms}}$ at the stress-free boundary of the fluid layer, we can define a standard Reynolds number, $Re=U_{\mathrm{rms}}L_f/\nu$. In the inertial regime with significant secondary flows, it is shown that $Re_F\delta^2\approx 2Re\delta^2$ \citep{FloresRamirez2025AsymmetricFlows}. In shallow coastal seas, for example, where $U_{\mathrm{rms}}\sim 0.1~{\rm{m}}/{\rm{s}}$, $L_f\sim 10^5~{\rm{m}}$, $\delta \sim 0.01$ \citep{Klein-2009} and the turbulent viscosity is approximately equal to the value used by \citet{Chen2021DielPatchiness}, we obtain $Re \delta^2 \sim 100$. This suggests that, while idealized, our results may represent regimes relevant for horizontal transport in certain environmental settings. Finally, it should be noted that in large-scale geophysical flows, rotation, which is neglected in our simulations and can be addressed in future studies, can further influence particle trajectories and the geometrical properties and spatial configuration of their distributions over depth. This can occur through mechanisms such as the presence of Ekman spirals in the boundary layers near the free surface and the bottom of the basin or the impact of Ekman pumping.} These elements should be included in more detail in the next steps to bridge the gap between the current model {\color{black}study, which can be used as a reference step, avoiding the complication of including rotation, to understand how certain transport phenomena are reflected in and can be understood with a few statistical metrics,} and applications in {\color{black}environmental and geophysical flows.}

This study was motivated by observational contexts in shallow lakes or coastal regions where only surface drifter data are typically available due to economic or technical constraints. Although our idealized setup does not allow direct extrapolation to real-world environments, our findings suggest that, even in shallow systems, the use of surface dispersion to reliably represent deeper dispersion is limited to the uppermost portion of the fluid layer. This limitation arises because key flow properties, such as horizontal flow divergence, can vary significantly with depth, shaping the spatial organization and accumulation of depth-keeping particles. For example, in the ocean, plastic waste patches on the surface become increasingly diffuse and eventually vanish at depth \cite{Wichmann2019InfluenceMicroplastic}. This is analogous to the breakdown of the particle filaments observed in our simulations at low $Re_F\delta^2$ (see Fig.~\ref{fig:particle_positions_ReF220_delta03}). This ``leakage" of plastics has important implications for understanding their large-scale transport from subtropical to polar regions. Consequently, to accurately infer deeper transport pathways from surface observations, one may require a detailed knowledge of the vertical structure of the horizontal velocity field.

\bmhead{Acknowledgements}

L.M.F.R. gratefully acknowledges financial support from the Consejo Nacional de Humanidades, Ciencias y Tecnolog\'ias (CONAHCYT, M\'exico) through a scholarship grant (No. 710021).

\section*{Declarations}

\begin{itemize}
\item Funding - L.M.F.R has received financial support from the Consejo Nacional de Humanidades, Ciencias y Tecnolog\'ias (CONAHCYT, M\'exico) through a scholarship grant (No. 710021) 
\item Conflict of interest/Competing interests - The authors have no competing interests to declare that are relevant to the content of this article.
\item Ethics approval and consent to participate - Not applicable
\item Consent for publication - Not applicable
\item Data availability - Data sets generated during the current study are available from the corresponding author on reasonable request.
\item Materials availability - Not applicable
\item Code availability - Not applicable
\item Author contribution - L.M.F.R. led the formal analysis and funding acquisition, and contributed to conceptualization, investigation, software, validation, visualization, and drafting the manuscript. M.D.M. was involved in conceptualization, investigation, supervision, co-wrote the original draft, and supported the review and analysis. H.J.H.C. contributed to conceptualization, investigation, project administration, supervision, and validation, and co-wrote the original draft and led the review and editing.
\end{itemize}

\noindent

\bibliography{references}

\end{document}